

{\def\unredoffs{\hoffset-.14truein\voffset-.21truein
                     \vsize=9.5truein}
\newbox\leftpage \newdimen\fullhsize \newdimen\hstitle \newdimen\hsbody
\tolerance=1000\hfuzz=2pt
\catcode`\@=11 
\def\bigans{b }
\magnification=1100\unredoffs\baselineskip=16pt plus 2pt minus 1pt
\hsbody=\hsize\hstitle=\hsize
\def\almostshipout#1{
\count1=1 \message{[\the\count0.\the\count1]}
      \global\setbox\leftpage=#1 \global\let\l@r=R}
\newcount\yearltd\yearltd=\year\advance\yearltd by -1900

%
%

\def\draftmode{\message{ DRAFTMODE }\def\draftdate{{\rm preliminary draft:
\number\month/\number\day/\number\yearltd\ \ \hourmin}}%
\headline={\hfil\draftdate}\writelabels\baselineskip=10pt plus 2pt minus 2pt
 {\count255=\time\divide\count255 by 60 \xdef\hourmin{\number\count255}
  \multiply\count255 by-60\advance\count255 by\time
  \xdef\hourmin{\hourmin:\ifnum\count255<10 0\fi\the\count255}}}
\def\nolabels{\def\wrlabeL##1{}\def\eqlabeL##1{}\def\reflabeL##1{}}
\def\writelabels{\def\wrlabeL##1{\leavevmode\vadjust{\rlap{\smash%
{\line{{\escapechar=` \hfill\rlap{\sevenrm\hskip.03in\string##1}}}}}}}%
\def\eqlabeL##1{{\escapechar-1\rlap{\sevenrm\hskip.05in\string##1}}}%
\def\reflabeL##1{\noexpand\llap{\noexpand\sevenrm\string\string\string##1}}}
\nolabels
%
\global\newcount\secno \global\secno=0
\global\newcount\meqno \global\meqno=1
\def\newsec#1{\global\advance\secno by1\message{(\the\secno. #1)}
\global\subsecno=0\eqnres@t\noindent{\bf\the\secno. #1}
\writetoca{{\secsym} {#1}}\par\nobreak\medskip\nobreak}
\def\eqnres@t{\xdef\secsym{\the\secno.}\global\meqno=1\bigbreak\bigskip}
\def\sequentialequations{\def\eqnres@t{\bigbreak}}\xdef\secsym{}
\global\newcount\subsecno \global\subsecno=0
\def\subsec#1{\global\advance\subsecno by1\message{(\secsym\the\subsecno. #1)}
\ifnum\lastpenalty>9000\else\bigbreak\fi
\noindent{\it\secsym\the\subsecno. #1}\writetoca{\string\quad
{\secsym\the\subsecno.} {#1}}\par\nobreak\medskip\nobreak}
\def\appendix#1#2{\global\meqno=1\global\subsecno=0\xdef\secsym{\hbox{#1.}}
\bigbreak\bigskip\noindent{\bf Appendix #1. #2}\message{(#1. #2)}
\writetoca{Appendix {#1.} {#2}}\par\nobreak\medskip\nobreak}
%
%
\def\eqnn#1{\xdef #1{(\secsym\the\meqno)}\writedef{#1\leftbracket#1}%
\global\advance\meqno by1\wrlabeL#1}
\def\eqna#1{\xdef #1##1{\hbox{$(\secsym\the\meqno##1)$}}
\writedef{#1\numbersign1\leftbracket#1{\numbersign1}}%
\global\advance\meqno by1\wrlabeL{#1$\{\}$}}
\def\eqn#1#2{\xdef #1{(\secsym\the\meqno)}\writedef{#1\leftbracket#1}%
\global\advance\meqno by1$$#2\eqno#1\eqlabeL#1$$}
%
\newskip\footskip\footskip14pt plus 1pt minus 1pt 
\def\footnotefont{\ninepoint}\def\f@t#1{\footnotefont #1\@foot}
\def\f@@t{\baselineskip\footskip\bgroup\footnotefont\aftergroup\@foot\let\next}
\setbox\strutbox=\hbox{\vrule height9.5pt depth4.5pt width0pt}
\global\newcount\ftno \global\ftno=0
\def\foot{\global\advance\ftno by1\nobreak\footnote{$^{\the\ftno}$}}
%
\newwrite\ftfile
\def\footend{\def\foot{\global\advance\ftno by1\chardef\wfile=\ftfile
$^{\the\ftno}$\ifnum\ftno=1\immediate\openout\ftfile=foots.tmp\fi%
\immediate\write\ftfile{\noexpand\smallskip%
\noexpand\item{f\the\ftno:\ }\pctsign}\findarg}%
\def\footatend{\vfill\eject\immediate\closeout\ftfile{\parindent=20pt
\centerline{\bf Footnotes}\nobreak\bigskip\input foots.tmp }}}
\def\footatend{}
%
%
\global\newcount\refno \global\refno=1
\newwrite\rfile
\def\ref{\nobreak[\the\refno]\nref}
\def\nref#1{\xdef#1{[\the\refno]}\writedef{#1\leftbracket#1}%
\ifnum\refno=1\immediate\openout\rfile=refs.tmp\fi
\global\advance\refno by1\chardef\wfile=\rfile\immediate
\write\rfile{\noexpand\item{#1\ }\reflabeL{#1\hskip.31in}\pctsign}\findarg}
\def\findarg#1#{\begingroup\obeylines\newlinechar=`\^^M\pass@rg}
{\obeylines\gdef\pass@rg#1{\writ@line\relax #1^^M\hbox{}^^M}%
\gdef\writ@line#1^^M{\expandafter\toks0\expandafter{\striprel@x #1}%
\edef\next{\the\toks0}\ifx\next\em@rk\let\next=\endgroup\else\ifx\next\empty%
\else\immediate\write\wfile{\the\toks0}\fi\let\next=\writ@line\fi\next\relax}}
\def\striprel@x#1{} \def\em@rk{\hbox{}}
\def\lref{\begingroup\obeylines\lr@f}
\def\lr@f#1#2{\gdef#1{\ref#1{#2}}\endgroup\unskip}
\def\semi{;\hfil\break}
\def\addref#1{\immediate\write\rfile{\noexpand\item{}#1}} 
\def\startrefs#1{\immediate\openout\rfile=refs.tmp\refno=#1}
\def\xref{\expandafter\xr@f}\def\xr@f[#1]{#1}
\def\refs#1{\count255=1\nobreak[\r@fs #1{\hbox{}}]}
\def\r@fs#1{\ifx\und@fined#1\message{reflabel \string#1 is undefined.}%
\nref#1{need to supply reference \string#1.}\fi%
\vphantom{\hphantom{#1}}\edef\next{#1}\ifx\next\em@rk\def\next{}%
\else\ifx\next#1\ifodd\count255\relax\xref#1\count255=0\fi%
\else#1\count255=1\fi\let\next=\r@fs\fi\next}
%

%
\newwrite\ffile\global\newcount\figno \global\figno=1
\def\fig{fig.~\the\figno\nfig}
\def\nfig#1{\xdef#1{fig.~\the\figno}%
\writedef{#1\leftbracket fig.\noexpand~\the\figno}%
\ifnum\figno=1\immediate\openout\ffile=figs.tmp\fi\chardef\wfile=\ffile%
\immediate\write\ffile{\noexpand\medskip\noexpand\item{Fig.\ \the\figno. }
\reflabeL{#1\hskip.55in}\pctsign}\global\advance\figno by1\findarg}
\def\xfig{\expandafter\xf@g}\def\xf@g fig.\penalty\@M\ {}
\def\figs#1{figs.~\f@gs #1{\hbox{}}}
\def\f@gs#1{\edef\next{#1}\ifx\next\em@rk\def\next{}\else
\ifx\next#1\xfig #1\else#1\fi\let\next=\f@gs\fi\next}
\newwrite\lfile
{\escapechar-1\xdef\pctsign{\string\%}\xdef\leftbracket{\string\{}
\xdef\rightbracket{\string\}}\xdef\numbersign{\string\#}}

\def\writestop{\def\writestoppt{\immediate\write\lfile{\string\pageno%
\the\pageno\string\startrefs\leftbracket\the\refno\rightbracket%
\string\def\string\secsym\leftbracket\secsym\rightbracket%
\string\secno\the\secno\string\meqno\the\meqno}\immediate\closeout\lfile}}
\def\writestoppt{}\def\writedef#1{}
\def\seclab#1{\xdef #1{\the\secno}\writedef{#1\leftbracket#1}\wrlabeL{#1=#1}}
\def\subseclab#1{\xdef #1{\secsym\the\subsecno}%
\writedef{#1\leftbracket#1}\wrlabeL{#1=#1}}
\newwrite\tfile \def\writetoca#1{}
\def\leaderfill{\leaders\hbox to 1em{\hss.\hss}\hfill}
\def\writetoc{\immediate\openout\tfile=toc.tmp
   \def\writetoca##1{{\edef\next{\write\tfile{\noindent ##1
   \string\leaderfill {\noexpand\number\pageno} \par}}\next}}}

%
\catcode`\@=12 
%
\edef\tfontsize{scaled\magstep3}
 \tfontsize  \tfontsize
 \tfontsize \font\titlei=cmmi10 \tfontsize
\font\titleis=cmmi7 \tfontsize \font\titleiss=cmmi5 \tfontsize
\font\titlesy=cmsy10 \tfontsize \font\titlesys=cmsy7 \tfontsize
\font\titlesyss=cmsy5 \tfontsize  \tfontsize
\skewchar\titlei='177 \skewchar\titleis='177 \skewchar\titleiss='177
\skewchar\titlesy='60 \skewchar\titlesys='60 \skewchar\titlesyss='60
\font\ninerm=cmr9 \font\sixrm=cmr6 \font\ninei=cmmi9 \font\sixi=cmmi6
\font\ninesy=cmsy9 \font\sixsy=cmsy6 \font\ninebf=cmbx9
\font\nineit=cmti9 \font\ninesl=cmsl9 \skewchar\ninei='177
\skewchar\sixi='177 \skewchar\ninesy='60 \skewchar\sixsy='60
\def\ninepoint{\def\rm{\fam0\ninerm}
\textfont0=\ninerm \scriptfont0=\sixrm \scriptscriptfont0=\fiverm
\textfont1=\ninei \scriptfont1=\sixi \scriptscriptfont1=\fivei
\textfont2=\ninesy \scriptfont2=\sixsy \scriptscriptfont2=\fivesy
\textfont\itfam=\ninei \def\it{\fam\itfam\nineit}\def\sl{\fam\slfam\ninesl}%
\textfont\bffam=\ninebf \def\bf{\fam\bffam\ninebf}\rm}
%
%
\def\noblackbox{\overfullrule=0pt}
\def\inv{^{\raise.15ex\hbox{${\scriptscriptstyle -}$}\kern-.05em 1}}

\def\Dsl{\,\raise.15ex\hbox{/}\mkern-13.5mu D} 
\def\dsl{\raise.15ex\hbox{/}\kern-.57em\partial}

\def\lspace{\ifx\answ\bigans{}\else\qquad\fi}
\def\lbspace{\ifx\answ\bigans{}\else\hskip-.2in\fi} 
\def\boxeqn#1{\vcenter{\vbox{\hrule\hbox{\vrule\kern3pt\vbox{\kern3pt
	\hbox{${\displaystyle #1}$}\kern3pt}\kern3pt\vrule}\hrule}}}
\def\mbox#1#2{\vcenter{\hrule \hbox{\vrule height#2in
		\kern#1in \vrule} \hrule}}  
\def\tilde{\widetilde} \def\bar{\overline} \def\hat{\widehat}


\pageno=0\nopagenumbers\tolerance=10000\hfuzz=5pt
\line{\hfill CERN-TH.7233/94}
\line{\hfill RAL-94-038}
\vskip 36pt
\centerline{\bf INFRARED AND ULTRAVIOLET BEHAVIOUR OF}
\centerline{\bf EFFECTIVE SCALAR FIELD THEORY}
\vskip 36pt\centerline{Richard~D.~Ball$^{a}$\footnote*{On leave of
absence from a Royal Society University Research Fellowship.}
and Robert~S.~Thorne$^{b}$}
\vskip 12pt
\centerline{\it Theory Division, CERN,}
\centerline{\it CH-1211 Gen\`eve 23, Switzerland.$^{a}$}
\vskip 10pt
\centerline{\it and }
\vskip 10pt
\centerline{\it Rutherford Appleton Laboratory,}
\centerline{\it Chilton, Didcot, Oxon., OX11 0QX, U.K.~$^{b}$}
\vskip 1.in
{\narrower\baselineskip 10pt
\centerline{\bf Abstract}
\medskip
We consider the infrared and ultraviolet behaviour of the effective
quantum field theory of a single $Z_2$ symmetric scalar field.
In a previous paper we proved to all orders in perturbation theory
the renormalizability of massive effective scalar field theory using
Wilson's exact renormalization group equation. Here we
show that away from exceptional momenta the massless theory is
similarly renormalizable, and we prove detailed bounds on Green's
functions as arbitrary combinations of exceptional Euclidean momenta
are approached. As a corollary we also prove
Weinberg's Theorem for the massive effective theory, in the form of
bounds on Green's functions at Euclidean momenta much greater than
the particle mass but below the naturalness scale of the theory.
}

\vskip 0.8in
\line{CERN-TH.7233/94\hfill}
\line{RAL-94-038\hfill}
\line{April 1994\hfill}
\vfill\eject
\footline={\hss\tenrm\folio\hss}


\def\frac#1#2{{{#1}\over {#2}}}
\def\half{\hbox{${1\over 2}$}}

\def\blackbox{\vrule height7pt width5pt depth2pt}

\def\VertL{\Vert_{\Lambda}}\def\VertR{\Vert_{\Lambda_R}}
\def\Real{\Re e}
\def\bk{\bar{k}}
\catcode`@=11 
\def\lsim{\mathrel{\mathpalette\@versim<}}
\def\gsim{\mathrel{\mathpalette\@versim>}}
 \def\@versim#1#2{\lower0.2ex\vbox{\baselineskip\z@skip\lineskip\z@skip
       \lineskiplimit\z@\ialign{$\m@th#1\hfil##$\crcr#2\crcr\sim\crcr}}}
\catcode`@=12 

\def\rhs{right hand side}
\def\PR{{\it Phys.~Rev.~}}

\def\NP{{\it Nucl.~Phys.~}}

\def\PRep{{\it Phys.~Rep.~}}

\def\CMP{{\it Comm.~Math.~Phys.~}}
\def\JMP{{\it Jour.~Math.~Phys.~}}

\def\vol#1{{\bf #1}}\def\vyp#1#2#3{\vol{#1} (#2) #3}

\noblackbox


\nref\REFT{R.~D.~Ball and R.~S.~Thorne, OUTP-93-23P, CERN-TH.7067/93,
to be published in {\it Ann.~Phys.}}
\nref\rWil{K.~G.~Wilson and J.~G.~Kogut, \PRep\vyp{12C}{1974}{75}.}
\nref\rPol{J. Polchinski, \NP\vyp{B231}{1984}{269}.}
In a previous paper \REFT\ we used the exact renormalization
group \refs{\rWil,\rPol} to construct a stable, unitarity and causal
effective massive scalar field
theory with a $Z_2$ global symmetry, for which we could then prove
perturbative renormalizability --- boundedness, convergence and
universality --- for processes at energy scales similar to the mass
of the particle but far below some naturalness scale. These results
were trivially extended to theories with arbitrary numbers of particles
of spin zero or one half, with linearly realized global symmetries.

However we only really considered theories with two different
scales, $\Lambda_R$ of order the masses of the particles, and
$\Lambda_0$, the naturalness scale, at which all the interactions
could become equally important and the effective theory becomes equivalent to
a general S--matrix theory. We might hope that we could be
more ambitious than this, since we often in practice wish to consider
situations where we have a number of different scales to be dealt with
simultaneously. An obvious example would be to have two particles with
significantly different masses; for processes at scales of order the
light particle mass one could then consider the extent to which the
effects of the heavy particle decouple. This is discussed in an
accompanying paper \ref\Decoupling{R.~D.~Ball and R.~S.~Thorne,
CERN-TH.????/94, RAL-94-038.}. In the present paper
we consider instead processes at
scales much larger than the mass scale of the theory. We first
consider a massless theory, proving bounds on Green's functions
corresponding to processes at finite scales $\Lambda_R\ll\Lambda_0$,
but with arbitrary combinations of (almost) exceptional momenta.
It is then relatively straightforward to turn the argument around, and
consider high energy processes of a massive theory (the only real
difference between the two cases being the renormalization conditions
on relevant couplings, as we will see). In this way both infrared and
ultraviolet behaviour may be examined within the same framework.

Within the conventional formulation of quantum field theory
the behaviour of a regulated Feynman graph when all external Euclidean
momenta are large was given by Weinberg's theorem
\ref\rwt{S.~Weinberg, \PR\vyp{118}{1960}{838}.}. Historically
Weinberg's theorem was the final step in the proof of perturbative
renormalizability of local quantum field theory. Similar techniques
may be used to deduce heuristically the infrared structure of
the unrenormalized theory when there are no exceptional momenta
\ref\rIRi{E.~C.~Poggio and H.~R.~Quinn, \PR\vyp {D14}{1976}{578}.}.
Much more detailed studies based on similar methods to those used
by Landau to find threshold singularities were made by Kinoshita
\ref\rIRii{T.~Kinoshita, \JMP\vyp{3}{1962}{650}
\semi T.~Kinoshita and A.~Ukawa, \PR\vyp{D13}{1976}{1573}.}, while
a rigorous but complicated analysis based on BPHZ subtraction was made
by Symanzik \ref\rIRSym{K.~Symanzik, \CMP\vyp{18}{1970}{227};
\vyp{23}{1071}{49}; \vyp{34}{1973}{7}.}. The main reason for this
technical complication of what should be, after all, simply a matter
of scaling, is that in the conventional formulation of perturbative
quantum field theory all the different scales in the theory come
into play at the same time, and ultraviolet and infrared problems
must be dealt with together. However, in the exact renormalization
group approach each scale is treated separately; in perturbation
theory the graphs are built up by successively nesting high momentum
loops in a strict hierarchy of decreasing scales. In this way
ultraviolet and infrared issues can be completely disentangled, and
the derivation of bounds on Green's functions in either the massless
limit or in the deep Euclidean region becomes relatively
straightforward.

We begin with a brief summary of the exact renormalization group, in
the formulation used in ref.\REFT. In \S 2 we discuss the
definition of a massless scalar theory and its infrared divergences,
show how to obtain bounds on the flow equations for arbitrary
combinations of exceptional momenta, use these to prove a set of
bounds on the renormalized Green's functions, and then use these to
discuss the infrared finiteness of physical S--matrix elements.
In \S 3 we derive a
similar set of bounds on the Green's functions of a massive effective
theory in the deep Euclidean region, which amount to an extension of
Weinberg's Theorem, and show further how these bounds may be
systematically improved.

\newsec{Effective Field Theory and the Exact Renormalization Group.}

An effective field theory may be defined through a classical Euclidean
action
\eqn\classact
{S[\phi;\Lambda] = \half\big(\phi,P\inv_{\Lambda}\phi\big)
+ S_{\rm int}[\phi;\Lambda].}
It is assumed that this action is Lorentz invariant, an
analytic functional of the fields $\phi$ and their derivatives,
and is natural at some scale $\Lambda_0$. The
propagator function is taken to be of the form
\eqn\exxix{P_{\Lambda}(p) = {K_{\Lambda}(p) \over (p^2 + m^2)},}
where $0<m^2\ll\Lambda_0^2$, and the regulating function
$K_{\Lambda}(p)$ is of the form
$K_{\Lambda}(p)\equiv K\big((p^2 + m^2)/\Lambda^2\big)$, with $K(x)$
real, positive, and monotonically decreasing in $x$,
$K(0)=1$, while both $K(z)$ and
$1/K(z)$ are regular functions, with an essential singularity at the
point at infinity. If the order of this singularity is sufficiently
large, the S--matrix of the theory will be finite.
The interaction may be expanded in the general form
\eqn\exxxi{S_{\rm int}[\phi;\Lambda] \equiv \sum^{\infty}_{m=1} \sum
^{\infty}_{r=1} {g^r\over (2m)!}
\int {d^4p_1\cdots d^4p_{2m} \over (2\pi)^{4(2m - 1)}}
V^r_{2m}(p_1,\ldots ,p_{2m};\Lambda)
\delta^4\big(\hbox{$\sum^{2m}_{i=1}p_i$}\big)
\phi_{p_1}\cdots\phi_{p_{2m}},}
where the vertex functions $V^r_{2m}(p_1,\ldots ,p_{2m};\Lambda)$ are,
for $\Lambda >0$, analytic functions of their arguments, and $g$ is a
coupling constant used to order the perturbation series. We may assume
that the theory has a $Z_2$ global symmetry, so that only even terms occur
in the expansion.

It is shown in \REFT\ that it is possible to construct actions of the
form \classact\ such that, despite the presence of derivatives of the
fields of arbitrarily high order, the classical equations of motion have only
the usual vacuum and free--particle solutions, allowing for the
construction of in-- and out--states, and thus of a perturbative
S--matrix. It is shown furthermore that this S--matrix is unitary and
causal at all scales.

The quantum theory may be defined by reducing $\Lambda$ from the
naturalness scale $\Lambda_0$ down to zero; the regulating function in
the propagator \exxix\ then ensures that all modes are successively
integrated over. The connected amputated Green's functions
$\tilde{G}^c_{2m}$ are invariant under changes of $\Lambda$ provided
the effective interaction satisfies the exact renormalization
group equation \refs{\rWil,\rPol}
\eqn\eiix
{{\partial S_{\rm int} \over \partial \Lambda}
= \half\int {d^4 p\over (2 \pi )^4}
{\partial P_{\Lambda}\over\partial\Lambda}
\biggl[{\delta S_{\rm int}\over\delta\phi_p}
       {\delta S_{\rm int}\over\delta\phi_{-p}}
     - {\delta^2 S_{\rm int}\over\delta\phi_p\delta\phi_{-p}} \biggr].}
So starting from a `bare' interaction $S_{\rm int}[\phi;\Lambda_0]$,
solution of \eiix\ yields a renormalized interaction
$S_{\rm int}[\phi;\Lambda]$; formally
\eqn\eix{\exp\bigl[-S_{\rm int}[\phi;\Lambda]-{\cal E}(\Lambda)\bigr]
=\exp ({\cal P}_{\Lambda_0}-{\cal P}_{\Lambda})
\exp -S_{\rm int}[\phi;\Lambda_0],}
where
\eqn\exi{{\cal P}_\Lambda\equiv\half\int{d^4 p\over(2\pi)^4}
P_{\Lambda}(p){\delta\over\delta\phi_p}{\delta\over\delta\phi_{-p}},}
and ${\cal E}(\Lambda)$ is a field independent constant. At
$\Lambda=0$ all modes have been integrated out and the amputated connected
Green's functions $\tilde{G}^c_{2m}$ may be read off from
$S_{\rm int}[\phi;0]$;
\eqn\eacgf{\tilde{G}^{c}_{2m}(p_1,\ldots ,p_{2m})\equiv
\prod_{i=1}^{2m}\Big(-\frac{\delta}{\delta \phi_{p_i}}\Big)
S_{\rm int}[\phi;0]\Big\vert_{\phi=0}.}

In terms of the vertex functions defined in \exxxi\ the
evolution equation \eiix\ becomes
\eqn\exxxxi
{\eqalign{\Lambda {\partial\over\partial \Lambda}
& V^r_{2m}(p_1\ldots,p_{2m};\Lambda)
=-{\Lambda^2 \over m(2m -1)}
\int {d^4p \over (2\pi\Lambda)^4}
K'_{\Lambda}(p) V^r_{2m+2}(p,-p,p_1\ldots,p_{2m};\Lambda) \cr
&\!\!\!+\sum^{m}_{l=1} \sum^{r-1}_{s=1}
{ \Lambda^{-2}2(2m)! \over (2m + 1 -2l)!(2l-1)!}
K'_{\Lambda}(P)
V^s_{2l}(p_1\ldots,p_{2l-1},P;\Lambda)V^{r-s}_{2m+2 -2l}
(-P,p_{2l}\ldots,p_{2m + 2};\Lambda),\cr}}
where $K'_{\Lambda}(p)\equiv K'\big((p^2+m^2)/\Lambda^2\big)$,
$K'(x)$ being the first derivative of the regulating function $K(x)$;
for a regulating function with an essential singularity of finite
order, $K'(x)=P(x)K(x)$ for some polynomial $P(x)$.
The amputated connected Green's functions are then
\eqn\evsc{\tilde{G}^c_{2m}
=\delta^4\big(\sum_{i=1}^{2m}p_i\big)\sum_r g^r V^r_{2m}(0),}
and S--matrix elements are given by analytic continuation and
addition of external lines. It can be shown that the S--matrix is independent
of the choice of regulating function, provided only that $K(z)$ is
chosen such that the
order of its essential singularity sufficiently large that the
integral in the first term on the \rhs\ of \exxxxi\ converges.

In \rPol\ and \REFT\ \exxxxi\ was used to obtain bounds on the vertex
functions and their derivatives; defining
\eqn\eNi{\Vert V\VertL \equiv
\max_{\{p_1,\ldots ,p_n\}}\biggl[\prod_{i=1}^{n}
[K_\Lambda(p_i)]^{1/4}
\vert V(p_1,...,p_{n};\Lambda)\vert\biggr],}
and using the simple inequalities
\eqn\exxxxiii{\eqalign{\int {d^4p \over (2\pi)^4}
\vert [K^{1/2}_{\Lambda}(p)] \vert &< C
\Lambda^4 \vert K^{1/2}(m^2/\Lambda^2)\vert,\cr
\bigl\vert K_{\Lambda}^{-1/2}(p)\partial^k_p K_{\Lambda}(p)
\bigr\vert &< D_k \Lambda^{-k}\vert K^{1/2}(m^2/\Lambda^2)\vert,\cr}}
we find for example that, ignoring constants on the \rhs,
\eqn\exxxxvii{\eqalign
{\Biggl\Vert{\partial \over \partial\Lambda} \biggl( \partial^j_p
V^r_{2m}(\Lambda) \biggr) \Biggr\Vert_{\bar\Lambda}
&\leq \bar\Lambda\Vert\partial^j_p V^r_{2m+2}(\Lambda)
\Vert_{\bar\Lambda} \cr
&\quad +\bar\Lambda^{-3}\sum^m_{l=1}
\sum^{r-1}_{s=1} \sum_{j_i;j_1 + j_2 + j_3 =j}
\bar\Lambda^{-j_1} \Vert \partial^{j_2}_p V^s_{2l}(\Lambda)
\Vert_{\bar\Lambda} \cdot
\Vert\partial^{j_3}_{p}V^{r-s}_{2m+2-2l}(\Lambda)\Vert_{\bar\Lambda},\cr}}
where $\bar\Lambda\equiv{\rm max}\{\Lambda,\Lambda_R\}$, $\Lambda_R$
being of order $m$. This bound is then used to prove a bound on the
amputated connected Green's functions \eacgf (and their derivatives);
to all orders in perturbation theory it turns out that
\eqn\elxxv{\Vert\partial^j_p \tilde G^{c}_{2m}\VertR
\leq  \Lambda_R^{4-2m-j}.}

\newsec{Infrared Behaviour}

\subsec{Defining a Massless Theory.}

To define a massless theory it is tempting simply to set $m=0$ in the
regularized propagator \exxix, and then further insist that at each
order in perturbation theory the renormalized
two--point vertex function vanishes at $\Lambda=0$ and at zero
momentum, ie. that $V^r_2(0,0;0)=0$. Clearly it would not be
sufficient to impose this condition at any finite renormalization
scale $\Lambda_R >0$, since then a mass term would be generated from
remaining evolution down to $\Lambda=0$, and the particle would not be
truly massless. It is thus imperative when dealing with massless
particles that we use a formulation of the exact renormalization group
(such as that described in \REFT, and outlined in the previous \S)
which allows the consideration of arbitrarily small regularization
scales $\Lambda$.\foot{In the formulation of the exact renormalization
group used in \rPol\ it is not possible to consider vertex functions
with external momenta above $\Lambda$; all low energy renormalization
scales must then be set at some finite scale $\Lambda_R$ and it is
impossible to consider theories with massless particles.}

However we know from experience that when calculating with massless
theories we encounter infrared divergences when partial sums of
momenta tend to zero. It is therefore necessary to see how these
divergences manifest themselves when using the exact renormalization
group. From \eix\ we can see that the vertices in the effective
action at some scale $\Lambda < \Lambda_0$ can be constructed from
diagrams consisting of the bare vertices and the regular `propagator'
$P_{\Lambda_0}(p)-P_{\Lambda}(p)$. Since there are manifestly no
infrared singularities in these diagrams the effective
action must be well--defined for all finite $\Lambda$. Indeed, this
statement is also true nonperturbatively; in the flow equation \eiix\
both terms are infrared finite even when $m=0$. Therefore,
we could try to regulate the theory in the infrared by defining a
minimum value of $\Lambda$,
$\Lambda_{\min}$, and then attempting to define the massless theory
by taking the limit $\Lambda_{\min} \rightarrow 0$. We might
then obtain a strictly massless theory if we set the
renormalization condition $V^r_{2}(P,-P;\Lambda_{\min}) \sim
\Lambda_{\min}^2$, where $\vert P \vert\sim\Lambda_{\min}$.

The problem with such an approach is that in the limit $\Lambda_{\min}\to
0$ the vertex functions $V^r_{2m}(p_i,\Lambda_{\min})$ remain regular
functions of momenta (or more specifically of the momentum invariants
$\{z_{ij}\equiv (p_i+p_j)^2; i=1,\ldots,2m, j=1,\ldots,i\}$), while
the connected amputated Green's functions must, when analytically
continued to the physical region, contain a complicated set of
overlapping poles and cuts at $z_{ij}=0$. Indeed the crucial relationship
\evsc\ between the vertex functions in the limit $\Lambda\to 0$ and
the amputated connected Green's functions cannot be established when
$m=0$ because in this limit the regulating function no longer vanishes
for all Euclidean momenta $p$, but instead has nonvanishing support at
$p^2=0$: if $\sigma$ is the order of the essential singularity at infinity,
\eqn\ereglimz{\lim_{\Lambda\rightarrow 0} K(p^2/\Lambda^2)=
\cases{0,&if $\Real\, p^2>0$;\cr
       1,&if $p^2=0$;\cr
       0,&if $\Real\, p^2<0$ and $\sigma$ is even;\cr
       \infty,&if $\Real\, p^2<0$ and $\sigma$ is odd.\cr}}
So the correspondence between S--matrix elements and the Euclidean
vertex functions is broken at precisely the points we wish to
consider, namely at exceptional momenta where the Green's
functions may be infrared divergent.

It is thus necessary to disentangle the infrared divergences at
exceptional momenta from the poles and cuts by choosing an infrared
regulator which works even as $\Lambda$ goes to zero.
The obvious choice is to keep a small non-zero mass term in the
propagator \exxix, but with $m$ much less than
the scale $\Lambda_R$ at which we want to investigate the physics.
The identification \evsc\ will then hold for all external momenta,
so we can then examine how the amputated connected Green's
functions depend on the ratio $(\Lambda_R/m)$ at exceptional momenta
of order $m$ or less, and thus deduce the infrared behaviour of the
theory by considering them in the limit $m \to 0$.

This choice of infrared regulator is by no means unique, but has
certain advantages: it makes the extension of the results on the
infrared structure of the massless theory to the consideration of
Green's functions at large momenta in the massive theory (i.e. to
Weinberg's theorem) very straightforward, and furthermore
the techniques used here can be applied directly to situations
where we have a number of particles with significantly different
masses (as is required for the treatment of decoupling in ref.\Decoupling).

To ensure that the theory is massless in the limit $m\to 0$ we must simply
set the renormalization conditions on the two point vertex at
$\Lambda=0$ such that it is of order $m^2$ for momenta
of order $m$. We thus take the renormalization conditions at
$\Lambda =0$ to be
\eqn\eIxv
{\eqalign{\lim_{\Lambda\to 0}V^r_2(\tilde P,-\tilde P;\Lambda) &=
\Lambda_m^2\hat\lambda^r_1,\cr
\lim_{\Lambda\to0}\bigl[
\partial_{p_{\mu}}\partial_{p_{\nu}}V^r_2(p,-p;\Lambda)
\vert_{p=P_0}\bigr]_{\delta_{\mu\nu}} &=\hat\lambda^r_2,\cr
\lim_{\Lambda\to0}V^r_4(P_1,P_2,P_3,P_4);\Lambda) &=
\hat\lambda^r_3.\cr}}
Here $\Lambda_m$ is some scale of order $m$, so
$\Lambda_m\ll\Lambda_R$, the renormalization scale,
$\tilde P$ and $P_i$ (such that $\sum_1^4 P_i=0$) are
the external momenta at which the renormalization
conditions are set, while $\hat \lambda^r_i$ are
some renormalization constants chosen independently of $\Lambda_0$,
$\Lambda_R$ and $\Lambda_m$. We choose $P_i$ and all their partial or
complete sums with magnitude similar to
$\Lambda_R$, while $\tilde P$ has magnitude similar to or less than
$\Lambda_m$.  Thus, except for the renormalization
condition on the two-point vertex, the renormalization conditions are
set for momenta much larger than the mass of the particle, and in
particular for non--vanishing momenta as $m \rightarrow 0$.\foot{It
would be possible to set the renormalization condition on the second
momentum derivative of the two point vertex at $P$ also. However,
this would lead to the necessity of a wave-function renormalization
which behaves like $P\log(\Lambda_R/\Lambda_m)$. As long as we were to
define our Green's functions via this wavefunction renormalization
before taking the limit $m\rightarrow 0$, we would still only obtain
infrared divergences at exceptional momenta, but we believe that the
bounding argument is far clearer if we avoid this complication.}

The renormalization conditions on irrelevant vertices at $\Lambda_0$
are (for convenience) taken to be the same as in \S 2 of \REFT, namely
\eqn\exxxv{\partial^j_p V^r_{2m}(\Lambda_0) = 0 \qquad 2m + j > 4.}
They will later be relaxed in order to prove the universality of
the massless theory.

\subsec{Exceptional Momenta.}

Now that the theory is well defined, we can begin to investigate its
infrared structure, by considering the behaviour of the vertices when
various combinations of their external momenta become small.
However, before we do this it will be convenient to
introduce some new terminology, in order to specify precisely what we
mean by this; it is not only the magnitudes of single momenta that
are important in general but that of the
sum of a set of momenta. Consider a particular
set $\{p_{\sigma(1)},\ldots,p_{\sigma(n)}\}$ of $n$ momenta (where
$\sigma(i)$ is some permutation of $i=1,\ldots,2m$). We define this
set\foot{
In order to prove perturbative renormalizability and infrared
finiteness of a massless scalar field theory using renormalization
group flow, it will be necessary to consider all possible combinations of
exceptional momenta. The proof of infrared finiteness in
ref.\ref\rIRiii{M. Bonini,
M. D'Attanasio and G. Marchesini, \NP\vyp{408}{1993}{441}.}, in which
momenta are only allowed to become exceptional in pairs, so that each
exceptional set contains only two elements, is thus
incomplete.} to be `exceptional' if the magnitude of the sum of
its momenta is less than a certain
value, which in practice we choose to be the scale at which the
renormalization conditions are set, $\Lambda_R$. In fact it will also
prove useful to impose a lower bound $E$ on the momenta, so that in fact
$E^2<(\sum_{i=1}^n p_{\sigma(i)})^2<\Lambda_R^2$.

For a vertex with $2m$ legs we always have at least one exceptional set
$\{p_1,\ldots,p_{2m}\}$ simply because
the sum of the momenta entering the
vertex is in practice equal to zero because of the delta function in
the definition \exxxi\ which enforces momentum conservation. We can
also see that because of momentum conservation
as soon as we have one exceptional set we automatically have two;
if there is an exceptional set $\{p_{\sigma(1)},\ldots,p_{\sigma(n)}\}$
containing $n$ momenta there is also another one
$\{p_{\sigma(2m-n+1)},\ldots,p_{\sigma(2m)}\}$ containing
$2m-n$ momenta.

We also wish to define a quantity $e$ which we may think of (loosely)
as the number of exceptional momenta for a given vertex; if we do this
appropriately then $e$ will actually turn out to give the degree of
infrared divergence at the vertex. To do this we first define an
irreducible exceptional set to be an exceptional set with no
exceptional subsets. We then may define the number of exceptional
momenta $e$ for a given vertex with $2m$ legs to be the total number
of distinct momenta
contained in the irreducible exceptional sets excluding those in the largest
irreducible exceptional set.\foot{This
definition means that $e$ is actually equal to the minimum value of
the total number of distinct momenta within exceptional sets
after the imposition of overall momentum conservation (which renders the
vertex a function of only $2m-1$ independent momenta).}
Note the the number of exceptional
momenta $e$ is not necessarily the same as the number of irreducible
exceptional sets; indeed it is not difficult to see that if the
largest irreducible exceptional set contains
$n$ momenta, then $e=2m-n$ irrespective of the details of the other
exceptional sets. Thus, for the
case where only the total sum of momenta is exceptional we would say that the
vertex has $e=0$; no exceptional momenta. If
there are only two exceptional sets, we must have the number of
exceptional momenta $e\leq m$, since if one set contains $n$
momenta, the other must contain $2m-n$. At the
other end of the scale, the maximum of value of $e$ is $2m-1$; this
can only occur if all the momenta are individually exceptional,
falling into $2m$ exceptional sets with one element each.

In order to illustrate this definition further we consider the simplest
cases. For the two point vertex the two external momenta sum to zero.
Thus for large values of $p^2$, we have a single irreducible
exceptional set of two momenta, $\{p,-p\}$, and
we take $e=0$; we have no exceptional momenta. If on the other hand
$p^2<\Lambda_R^2$, then we have two irreducible sets of
exceptional momenta, $\{p\}$ and $\{-p\}$, and we take $e=1$.
The possible combinations of irreducible sets (up to permutations) and
the corresponding number of exceptional momenta for the four point
vertex are:
$$\vbox{\settabs\+ &++++++++++++++++++++++&\cr
\+&$\{p_1,p_2,p_3,p_4\}$\hfill          &$e=0$,\cr
\+&$\{p_1,p_2,p_3\}\{p_4\}$\hfill       &$e=1$,\cr
\+&$\{p_1,p_2\}\{p_3,p_4\}$\hfill       &$e=2$,\cr
\+&$\{p_1,p_2\}\{p_3\}\{p_4\}$\hfill    &$e=2$,\cr
\+&$\{p_1,p_2\}\{p_3,p_4\}\{p_2,p_3\}\{p_1,p_4\}$\hfill&$e=2$,\cr
\+&$\{p_1\}\{p_2\}\{p_3\}\{p_4\}$\hfill &$e=3$.\cr}
$$

\subsec{Bounding the Flow Equations.}

We now consider the bounding of the vertices. As in \S 2.2 of \REFT\
we construct a norm $\Vert V \Vert_{\Lambda}$  of a vertex function
$V(p_1,...,p_{n})$ by multiplying each leg of the
vertex by $[K_\Lambda(p_i)]^{1/4}$ and finding the
maximum with respect to all the momenta, as in
\eNi. However, in order to investigate the dependence of the vertices
on momenta with magnitude less than $\Lambda_R$ we now also introduce the
idea of a restricted norm in which we find the maximum over a
restricted range of momenta. In practice we consider taking the
maximum over sets of momenta $\Pi(n,e;\Lambda_R,E)$ defined as the
set $\{p_1,\ldots,p_n\}$ such that there are $e$ exceptional momenta,
the total momentum in each exceptional set (including the largest) being
constrained to lie in the range $(E,\Lambda_R)$. The restricted norm
is thus
\eqn\eIix{\Vert V_{2m}(\Lambda)\Vert_{\Lambda_R}^{E,e} \equiv
\max_{\Pi(n,e;\Lambda_R,E)}\biggl[\prod_{i=1}^{n}
[K_{\Lambda_R}(p_i)]^{1/4}
\vert V_{2m}(p_1,...,p_{n};\Lambda)\vert\biggr].}
It is not difficult to see that this is indeed a norm.
If $e=0$ we will omit the superscripts $E$ and $e$.

We can use this definition of a restricted norm to bound both sides
of the flow equations \exxxxi. However, in order to do this
effectively we must also suitably extend the inequalities \exxxxiii.
The first of these is only used for bounding integrals over internal loop
momenta, independent of all external momenta, and thus does not
change if we put restrictions on the external momenta. The argument
$p$ in the second inequality \exxxxiii\ is however to be identified
with  $P$ in \exxxxi, which is a sum of external momenta. If we
restrict the form of the external momenta, we therefore restrict
the possible values of $P$, and thus require an improved form of this
inequality for such restricted $p$.
In fact we want to consider the left--hand side of this inequality
as a maximum over a range of $p$ greater than a minimum value,
$E\in [0,\Lambda_R)$. Remembering the form of $K_{\Lambda}(p)$, as
described following \exxix, and in particular its dependence on the
mass $m$, it is not difficult to show that
\eqn\eIxv{\max_{p > E}\Bigl\vert \Lambda^{-n}
K_{\Lambda}^{-1/2}(p)\partial^j_p K_{\Lambda}(p)
\Bigr\vert \leq \bar \Lambda^{-n-j},}
where $\bar \Lambda =\max(\Lambda_m,\Lambda,E)$, and $n$ is a
non--negative integer.

We are now nearly ready to bound the left--hand side of \exxxxi, or more
precisely, the left--hand side of \exxxxi\ after it has been differentiated
$j$ times with respect to external momenta. For
$\Lambda \in [\Lambda_R,\Lambda_0]$ we proceed in exactly the same way
as in \S 2 of \REFT; we take the norm with respect to $\Lambda$ and
do not concern ourselves with the values of the external momenta, to
give the bounded flow equation \exxxxvii\ with $\bar\Lambda=\Lambda$.

In the range $\Lambda\in [0,\Lambda_R]$ we must be rather more careful in our
bounding. In this case we consider the number of sets of momenta with
magnitudes as low as $E\in [0,\Lambda_R)$. (In fact, as in \eIxv, $E$ has an
effective lower cut--off of $\Lambda$ if $\Lambda>\Lambda_m$ and
$\Lambda_m$ if $\Lambda \leq \Lambda_m$,so it is only really necessary
to consider $E\in [{\rm max}(\Lambda,\Lambda_m),\Lambda_R)$ as we will
see shortly.) To this end we need to consider what type of norms we
have on the right--hand side of the
$j_{th}$ momentum derivative of \exxxxi\ once we have specified the type of
norm on the left--hand side (i.e. the number of exceptional momenta in
contains).

We first consider the first term on the right--hand side of
\exxxxi. If ${p_1,\ldots,p_{2m}}$ contain $e$
exceptional momenta then
$\partial^j_p V_{2m+2}(p,-p,p_1\ldots p_{2m};\Lambda)$ contains
exactly $e+2$ exceptional momenta. The
extra two exceptional momenta are only due to the fact that $p$ and
$-p$ sum to zero, and therefore obviously form a set of two which is
exceptional. ($p$ and $-p$ can take all
values and can therefore also form two exceptional sets of one. We
will see that this creates no new effects.) We might think that
since $p$ can take any value, then whenever $p$ is equal to the sum of some
set of momenta, and thus $p_{i_1} + \ldots p_{i_k} -p =0$, we would
have another exceptional configuration. This is strictly true, but rather
misleading: since
$V_{2m+2}(p,-p,p_1\ldots p_{2m};\Lambda)$ is invariant not only under
permutations of momenta, but also under $p\rightarrow -p$, the vertex
can only depend on $p$ through the invariants $p^2$ or
$((p_{i_1} + \ldots +p_{i_k}) \cdot p)^2$, and not on $(p_{i_i} +
\ldots + p_{i_k} \pm p)^2$. Thus, since the vertex does not depend on
$p_{i_1} + \ldots p_{i_k} -p$, its value has no significance and we
do not class it as an exceptional set.

The second term on the right--hand side of the $j_{th}$ momentum
derivative of \exxxxi\ is a little more difficult to deal with.
We first consider the $\Lambda$--derivative of the propagator which
links the two vertices in this term. The argument of this propagator
is $P= p_1+\ldots+p_{2l-1}$, which is the sum of all the external momenta on
either of the two vertices. In order to obtain the terms with the
largest value on the right--hand side of the $j_{th}$ momentum
derivative once we take the restricted norms, we must consider
an exceptional value of $P$. This is because when taking norms of
both sides of \exxxxi\ we will obtain a factor of $\bar \Lambda^{-3}$ from
$\max_{p\leq E}\vert\Lambda^{-3}K^{-{1 \over
2}}_{\Lambda}(p)\partial^jK_{\Lambda}(p)\vert$ if we have exceptional $P$,
rather than the
smaller value of $\Lambda_R^{-3}$ we would obtain for non--exceptional
$P$ (as we see from \eIxv). This exceptional value of $P$ is obtained
if all the external momenta on
each vertex comprise an
exceptional set, since all the external momenta
for the vertices will then sum to give $P$ with magnitude less than or equal
to $\Lambda_R$. This requirement is equivalent to demanding that none
of the irreducible sets of exceptional momenta for $V^r_{2m}$,
including the largest,
are split between the two vertices.

We now consider the number of exceptional momenta we may obtain for each of the
vertices in the second term on the right of \exxxxi\ for a given
number on the left--hand side. Since the bounds on the restricted
norms become larger for larger numbers of exceptional momenta (as we
will soon discover), in order to find the dominant terms on the right--hand
side
we need to have the largest value of the sum of the number of
exceptional momenta on each of the vertices that we can. We first consider
the case of exceptional $P$. In this case no sets of exceptional
momenta are split between the two vertices. Therefore, the largest set
of exceptional momenta for the vertex on the left--hand side resides
on one of the vertices in the term on the right--hand side. This
largest set contains $2m-e$ momenta, and must necessarily be the
largest set on the vertex on which it now resides. The largest set on
the other vertex on the right--hand side must contain at least one
momentum. Therefore, the largest sets on each vertex contain at least
$2m-e +1$ momenta between them. The two vertices have in all $2m+2$
legs, so the maximum value of $e_1 + e_2$, the sum of the number of
exceptional momenta on each vertex, is $e+1$. If $P$ is
non--exceptional then some of the exceptional sets on the left hand
side must be split between the two vertices on the right--hand side.
If the largest set is not split, then the largest set on one vertex
must contain $2m-e$ momenta. The largest set on the other vertex must
contain at least 2 momenta (else none of the original sets could have
been split), and using the same argument as above, the
maximum value of $e_1+e_2$ is $e$. If the largest set is split so that
$n\leq2m-e$ momenta go on one vertex and the remainder on another
vertex, then the largest set on one vertex must contain at least $n+1$
momenta, and the largest set on the other vertex must contain at least
$2m-e-n+1$ momenta. The maximum value of $e_1 +e_2$ is then $e$ again.

Finally, we must consider how momentum derivatives acting
on the left--hand side affects the right--hand side of
the bounded flow equation. The effect can occur in two ways: the
momentum derivative may act on one of the two vertices
in the second term on the right of \exxxxi, or may act on the
propagator linking these two vertices. As in the case of the
$\Lambda$-derivative of the propagator, we clearly obtain the largest
factors when bounding the momentum derivatives of the
$\Lambda$-derivative of the propagator when $P$ is exceptional, in
which case each momentum derivative gives a factor of $\bar
\Lambda^{-1}$, rather than $\Lambda_R^{-1}$ for unexceptional $P$.
Thus, the criterion of not splitting the sets of exceptional momenta
between the two vertices on the right--hand side leads to the largest factors
for the $\Lambda$-derivative of the propagator linking
the two vertices, whether there are momentum derivatives or not, and
also to the largest value of $e_1+e_2$.

\medskip

These results now allow us to obtain bounded flow equations that are useful
for bounding the vertices in ranges $\Lambda \in
[0,\Lambda_R]$. Considering all
$\Lambda\in [0,\Lambda_R]$,
we may act with $j$ momentum derivatives on the flow
equation \exxxxi, then take the norms, and using
\exxxxiii\ and \eIxv\ to bound the left--hand side, we obtain
\eqn\eIxi{\eqalign{\Biggl\Vert{\partial \over \partial\Lambda} \biggl(
\partial^{j}_p V^r_{2m}(\Lambda) \biggr) \Biggr\Vert_{\Lambda_R}^{E,e}
& \leq  \Bigl( \Lambda \,\Vert \partial^{j}_p V^r_{2m+2}(\Lambda)
\Vert_{\Lambda_R}^{E,e;0,2} \cr
&\hskip -1.35in + \sum_{l} \sum^{r-1}_{s=1} \sum_{j_i;j_1 + j_2 + j_3 =
j}\!\Bigl[ \bar \Lambda^{-3-j_{1}}(1 -
\delta_{l1}(\delta_{j_{2}1}+\delta_{j_{2}0})-
\delta_{(m+1-l)1}(\delta_{j_{3}1}+\delta_{j_{3}0})) \Vert \partial^{j_{2}}_p
V^s_{2l}(\Lambda) \Vert_{\Lambda_R}^{E,e_1} \cr
& \hskip 1.5 in\times \Vert
\partial^{j_{3}}_{p}  V^{r-s}_{2m + 2
-2l}(\Lambda) \Vert_{\Lambda_R}^{E,e_2}  \cr
&\hskip -1.35in +
(\delta_{l1}(\delta_{j_{2}1}+\delta_{j_{2}0})
+\delta_{(m+1-l)1}(\delta_{j_{3}1}+\delta_{j_{3}0}))
\sum_i \max_{p_i\geq E} \Bigl(\Lambda^{-3}
\vert\partial^{j_{1}}K_{\Lambda}(p_i)\vert\cdot\vert \partial^{j_{2}}_p
V^{s}_{2}(p_i;\Lambda)\vert\Bigr)\cr
&\hskip 1.5in\times
\Vert\partial^{j_{3}}_pV^{r-s}_{2m}(\Lambda)\Vert_{\Lambda_R}^{E,e}\Bigr]\Bigr)
+ \quad \hbox{other terms,}\cr}}
for $e\geq 1$, and $m>1$ or $m=1$, $j>1$.
The terms on the right--hand side with two vertices which are
written explicitly are those where the exceptional sets of momenta are not
split between the two vertices, and where at the maximum $e_1
+e_2=e+1$ (this maximum being reached automatically if we have a two
point vertex). The sum over $l$ does not run from from $1$ to $m$ as
it does in \exxxxvii\ because for a given
configuration of sets of exceptional momenta not all these values of
$l$ may be consistent with the requirement that no sets of exceptional
momenta are split.  The special terms for $m=1$, $j_{2}<2$ occur
because these vertices have rather special bounds. We already have some hint of
this in the fact that the
renormalization condition for $V^r_2$ is rather special. These special
terms will be simplified later. The
`other terms' correspond to ways of putting together two vertices such that
at least one set of exceptional momenta is split; such terms, as we
will prove, make smaller contributions to
the bound than the terms written explicitly. The form of the superscript on
the norm for the first term on the right of this equation is a little
different to the other superscripts. It signifies that since, as
we have already mentioned, this term has two momenta which may each
become as low as zero we denote it as above. We will discuss this
term in more detail later.

For the special case $m=1$, $j =0$, $p_1\leq \Lambda_R$, we have a slightly
different
type of inequality. (It is not necessary to derive an inequality for
$\partial_{p_{\mu}}V^r_2(\Lambda)$ since it can be entirely constructed from
$\partial^2_pV^r_2(\Lambda)$ using the Taylor formula about $p=0$.)
This time we simply take the maximum of the modulus of both sides of
\exxxxi\ obtaining
\eqn\eIiv{\eqalign{\max_{p\leq E} \Biggl\vert {\partial \over
\partial\Lambda} \biggl(
V^r_{2}(p;\Lambda) \biggr) \Biggr\vert
\leq  \max_{p\leq E}&\Bigl\vert \int d^4p'
\Lambda^{-3}K'_{\Lambda}(p)V^r_{4}(p',-p',p,-p;\Lambda)
\Bigr\vert\cr
&+\quad \sum_{s=1}^{r-1}\max_{p\leq E}\Bigl\vert \Lambda^{-3}
K'_{\Lambda}(p)V^s_{2}(p;\Lambda) V^{r-s}_{2}(p;\Lambda) \Bigr\vert.\cr}}
Since for $p\leq E$ factors of $K^{-1/4}_{\Lambda_R}(p)\sim 1$,
and thus the first term on the right--hand side of \eIiv\ is $\leq
\Lambda \Vert V^r_{4}(\Lambda)\Vert_{\Lambda_R}^{0,3}$ and
we may write \eIiv\ as
\eqn\eIv{\max_{p\leq E} \Biggl \vert {\partial \over \partial\Lambda} \biggl(
V^r_{2}(p;\Lambda) \biggr) \Biggr\vert
\leq  \Lambda \Vert V^r_{4}(\Lambda)\Vert_{\Lambda_R}^{0,3}
+\sum_{s=1}^{r-1} \max_{p\leq E}\Bigl\vert \Lambda^{-3}
K'_{\Lambda}(p)V^s_{2}(p;\Lambda)
V^{r-s}_{2}(p;\Lambda) \Bigr\vert.}

If $e=0$ then bounding the left hand side of \exxxxi\ is of course
much easier, since there
are then no exceptional momenta on either side of the flow equation, except
for the pair $\{p\}\{-p\}$ in the term $\partial^j_p
V^r_{2m+2}(p,-p,p_1\ldots p_{2m};\Lambda)$.
Taking norms we therefore obtain
\eqn\eIxii{\eqalign
{\Biggl \Vert {\partial \over \partial\Lambda} \biggl(
\partial^j_p V^r_{2m}(\Lambda) \biggr) \Biggr\Vert_{\Lambda_R}
&\!\!\leq  \Bigl( \Lambda\,\Vert \partial^j_p V^r_{2m+2}(\Lambda)
\Vert_{\Lambda_R}^{0,2} \cr
&\quad +\sum^m_{l=1} \sum^{r-1}_{s=1} \sum_{j_i;j_1 + j_2 + j_3 =
j} \hskip -0.15in \Lambda_R^{-3-j_1} \Vert \partial^{j_2}_p V^s_{2l}(\Lambda)
\Vert_{\Lambda_R} \cdot \Vert
\partial^{j_3}_{p} V^{r-s}_{2m + 2 -2l}(\Lambda) \Vert_{\Lambda_R} \Bigr).\cr}}

\subsec{Boundedness}

These equations together with the boundary conditions on the relevant
couplings, \eIxv, and the trivial boundary conditions \exxxv\ on the
irrelevant couplings, are all we need to prove\foot{To avoid confusion
the lemmas in this paper are numbered sequentially with those of our
previous paper \REFT.}
\medskip
\noindent{\it Lemma 5:}

\noindent i) For all $\Lambda \in [\Lambda_R,\Lambda_0]$,
\eqn\eIi{ \bigl\Vert \partial^j_p V^r_{2m}(\Lambda) \bigr\Vert_{\Lambda} \leq
\Lambda^{4-2m-j} \Biggl( P\log \biggl( {\Lambda \over \Lambda_R} \biggr)
+ {\Lambda \over\Lambda_0} P\log \biggl({\Lambda_0 \over
\Lambda_R}\biggr)\Biggr).}
\noindent ii) For all $\Lambda \in [0,\Lambda_R]$, and for
$e=0$, we have
\eqn\eIii{ \bigl\Vert \partial^j_p V^r_{2m}(\Lambda) \bigr\Vert_{\Lambda_R}
\leq
\Lambda_R^{4-2m-j}.}
\noindent iii) For all $\Lambda \in [0,\Lambda_R]$, and $1\leq e\leq
2m-3$, $m>1$ or $m=1$, $j>1$
\eqn\eIiii{ \bigl\Vert \partial^{j}_p  V^r_{2m}(\Lambda)
\bigr\Vert_{\Lambda_R}^{E,e} \leq\cases {
 \Lambda_R^{4-2m-j}
 \Bigl({\Lambda_R\over \bar\Lambda}\Bigr)^{e+j-1}
 P\log \Bigl({\Lambda_R \over \bar\Lambda}\Bigr) & $e$ odd,\cr
 \Lambda_R^{4-2m-j}
 \Bigl({\Lambda_R\over \bar\Lambda}\Bigr)^{e+j-2}
 P\log \Bigl({\Lambda_R \over \bar\Lambda}\Bigr) & $e$ even.\cr}}
For $e>2m-3$ the bound is simply
$\bar\Lambda^{4-2m-j} P\log ({\Lambda_R \over \bar\Lambda})$.

\noindent iv) The case $\Lambda \in [0,\Lambda_R]$, $e=1$, $m=1$ and
$j<2$, is a little different, since we do not have a restricted norm,
but an inequality reflecting the inequality \eIiv:
\eqn\eIxiix{ \max_{p \leq E}\bigl\vert \partial^j_p V^r_{2}(\Lambda)
\bigr\vert \leq
\bar\Lambda^{2-j} P\log \Bigl({\Lambda_R\over \bar\Lambda}\Bigr) }

\medskip

As for lemmas 1-4 of \REFT, the method of proof is induction, and the induction
scheme is
exactly the same as for the proof of lemma 1. Again we assume that the
lemma are true up to order $r-1$ in the expansion coefficient $g$, and
that at order $r$ in $g$ they true down to $m+1$.

\medskip

a) We first consider the irrelevant vertices for $\Lambda\in
[\Lambda_R,\Lambda_0]$. This step is identical to the first step in
the proof of lemma 1 in \S 2.2 of \REFT, since the flow equation,
the lemma and the boundary
conditions on the irrelevant vertices are exactly the same as they
were there. Thus, we verify i) simply by
using the flow equation and integrating from $\Lambda$ up to $\Lambda_0$.

\medskip

b) When considering the irrelevant vertices for $\Lambda
\in[0,\Lambda_R]$ we have to be more careful about the number of
exceptional momenta in external legs. We
first consider iii) for $\Lambda > \Lambda_m$, with $e$
odd and $\leq 2m-3$.
Integrating from $\Lambda$ up to $\Lambda_R$, taking norms, and using the
derived
boundary condition obtained by evaluating \eIi\ at $\Lambda_R$ we obtain
\eqn\eIxix{ \Vert
\partial^{j}_p V^r_{2m}(\Lambda)
\Vert_{\Lambda_R}^{E,e}\leq \Lambda_R^{4-2m-j} +
\int_{\Lambda}^{\Lambda_R}d\Lambda' \Biggl \Vert{\partial \over
\partial\Lambda'}  \biggl(
\partial^{j}_p  V^r_{2m}(\Lambda') \biggr)
\Biggr\Vert_{\Lambda_R}^{E,e}.}
We are able to bound the integrand in the second term on the right using the
bounds we already have on vertices at
orders $r-1$ and less in $g$ and vertices of $m+1$ or more legs, i.e.
those in ii), iii) and iv). However, before we
do this it is necessary to use the lemmas in order to simplify
\eIxi\ a little.

To begin we may use iv) to simplify the terms containing
$V^{s}_{2}(\Lambda)$
in \eIxi. We first consider $E\geq \Lambda$. From ii) we
know that for $p\geq
\Lambda_R$, $\partial^{j_{2}}_p V^{s}_2(\Lambda) \leq
K^{-1/2}_{\Lambda_R}(p)\Lambda_R^{-j_{2}}$, which grows less quickly than
$\partial^{j_1}_pK_{\Lambda}(p)$ falls. Hence, up to the usual
multiplicative constant, the maximum value of $(\Lambda^{-3}
\vert\partial^{j_{1}}K_{\Lambda}(p_i)\vert\cdot\vert \partial^{j_{2}}_p
V^{s}_{2}(p_i;\Lambda)\vert)$ in the range
$p\geq\Lambda_R$ is the value at $\Lambda_R$, i.e.
$\Lambda_R^{-1-j_1-j_{2}}$. For $p=E'$ where $E\leq E'\leq \Lambda_R$,
from iv) we know that $\vert \partial^{j_{2}}_pV^{s}_{2}(\Lambda)\vert\leq
(E')^{2-j_{2}}P\log(\Lambda_R/E')$, while
$\Lambda^{-3}\vert\partial^{j_{1}}_pK_{\Lambda}(p)\vert \leq
(E')^{-3-j_{1}}$ as we see from \eIxv. So the combined effect is
that $(\Lambda^{-3} \vert\partial^{j_{1}}K_{\Lambda}(p_i)\vert\cdot\vert
\partial^{j_{2}}_p
V^{s}_{2}(p_i;\Lambda)\vert)\leq (E')^{-1-j_{1}-j_{2}}P\log(\Lambda_R/E')$ in
this
range. Therefore, whether $j_{2}$ is zero or one, and whatever the value of
$j_{1}$, the maximum value of $(\Lambda^{-3}
\vert\partial^{j_{1}}K_{\Lambda}(p_i)\vert\cdot\vert \partial^{j_{2}}_p
V^{s}_{2}(p_i;\Lambda)\vert)$ for $p_1\geq E$ occurs when $p_1 =E$ and
is equal to $E^{-1-j_{1}-j_{2}}P\log(\Lambda_R/E)$. In this case we have the
normal
factor of
$E^{-3-j_{1}}$, as we do for the other explicit terms with two
vertices in \eIxi, and the two--point vertex and its first derivative
behaves as though it has a bound $E^{2-j_{2}}P\log(\Lambda_R/E)$,
i.e the same type of bound as its higher derivatives. The terms
on the right--hand side of \eIxi\ with the two--point function and its
first derivative having exceptional momenta thus behave like the other
explicitly written terms with two vertices. For $E<\Lambda$ the bound
on $(\Lambda^{-3}
\vert\partial^{j_{1}}K_{\Lambda}(p_i)\vert\cdot\vert \partial^{j_{2}}_p
V^{s}_{2}(p_i;\Lambda)\vert)$ is unchanged below $\Lambda$, being
equal to $\Lambda^{-1-j_1-j_2}P\log(\Lambda_R/\Lambda)$, which is
larger than that for all $E'>\Lambda$. This therefore gives us our maximum, and
when $E \leq \Lambda$ we obtain the normal factor of $\Lambda^{-3-j_1}$ as the
other
explicit terms with two vertices in \eIxi, i.e.
the result for the last term in \eIxi\ generalizes in the same way as
for $E\geq \Lambda$. This is clearly also true when both $E$ and
$\Lambda$ are $\leq \Lambda_m$.

We also consider the first term on the right of \eIxi, i.e. the bound
where the two internal momenta are equal and opposite. Since the growth of the
vertices for small momenta depends only on the number of exceptional
momenta, it is only the value of the arguments $\sum_{i}p_i$ of the
vertices
that determine this infrared behaviour. In general, a given
vertex $V^r_{2m+2}(p_a,p_b,p_1,\ldots,p_{2m};\Lambda)$ will depend on
all possible partial sums of momenta. However, as we
have already stated, if two of the momenta sum to zero, i.e
$p_a=-p_b=p$, then invariance under
permutations of the momenta guarantees there is no dependence on
$(\pm p+p_{i_1}+\ldots +p_{i_k})$. Imposing $p_a+p_b=0$ (and thus
$\sum_{i=1}^{2m}p_i=0$),
while keeping all subsets non--exceptional, we induce an infrared
factor of $P\log({\Lambda_R\over \Lambda})$ because we have a sum of
two momenta which is
exceptional with a value lower than $\Lambda$, in fact a value of
zero. (Of course, if $\Lambda\leq \Lambda_m$ we replace $\Lambda$ by
$\Lambda_m$ in the above.) From iii) this behaviour will not change
if we allow $p$ to become small on its own; we still have only two
exceptional momenta. The vertex is now a
function of $\Lambda$, $\Lambda_R$ and $\{p_1,\ldots,p_{2m}\}$ (and
$\Lambda_0$, to inverse powers, of course), and we can only obtain
the further infrared behaviour by letting partial sums of
$\{p_1,\ldots,p_{2m}\}$ become small, and this further infrared behaviour must
therefore depend only on the value of these partial sums. Thus, if we now let
subsets of
 $\{p_1,\ldots,p_{2m}\}$  become exceptional as low as
$E>\Lambda$, any new infrared behaviour can only depend on the
value of these subsets, i.e. on $E$. Similarly, any derivatives with
respect to $p_i \in \{p_1,\ldots,p_{2m}\}$ can only bring about
infrared behaviour depending on $E$.

We must verify the form of this $E$-dependent behaviour. We could consider
increasing $p_a$, $p_b$ and $p_a +p_b$ to a value $\sim E$, clearly having to
alter other
momenta do so; but only by order $E$. It
is obviously possible to do this in such a way that the number and
type of subsets of $\{p_1,\ldots,p_{2m}\}$ exceptional with respect to
$E$ is unchanged, and such that no partial sum of
$\{p_a,p_b,p_1,\ldots,p_{2m}\}$ sums to less than $E$. When we set
$p_a=-p_b$, the terms in
the expression $V^r_{2m+2}(p_a,p_b,p_1,\ldots ,p_{2m};\Lambda)$
involving partial sums including some of both $\{p_a,p_b\}$ and
$\{p_1,\ldots,p_{2m}\}$ all reduced to partial sums involving just the
$\{p_1,\ldots,p_{2m}\}$. Removing this
restriction on $p_a$ and $p_b$, as described above, we do not
qualitatively alter the value of any of these partial sums; if they
were exceptional as low as $E$ then they remain so, if not, then
changing their value by $\sim E$ does not make them so. Similarly, all
partial sums involving the $\{p_1,\ldots,p_{2m}\}$ in both cases
remain of the same order of magnitude. Thus, all the infrared
behaviour which was exhibited by
$V^r_{2m+2}(p,-p,p_1,\ldots,p_{2m};\Lambda)$
must remain
qualitatively the same when we make the type of change of momenta
outlined above, except that the logarithmic term in $(\Lambda_R/\Lambda)$ will
necessarily become a logarithmic term in $(\Lambda_R/E)$. The whole
must now satisfy $\Vert V^r_{2m+2}(\Lambda')\Vert_{\Lambda_R}^{E,e+2}$,
and in this way we see
that the powerlike $E$--dependence of
$V^r_{2m+2}(p,-p,p_1,\ldots,p_{2m};\Lambda)$ is the same as that for
$V^r_{2m+2}(p_a,p_b,p_1,\ldots ,p_{2m};\Lambda')$ in the infrared
region, and thus, from iii) we must have the bound
$\Vert V^r_{2m+2}(\Lambda)\Vert_{\Lambda_R}^{E,e;0,2} \leq
\Lambda_R^{3+e-2m}E^{-1-e-j}P\log({\Lambda_R\over \Lambda})$ for odd
$e$ (with obvious alteration for even $e$). Hence, the bound on the
first term on
the right--hand side of \eIxi\ will be $\Lambda\,
\Lambda_R^{3+e-2m}E^{-1-e-j}P\log({\Lambda_R\over \Lambda})$.\foot{
It would be possible to verify
this result in a rather direct manner by deriving bounds for vertices where
we let some sets of momenta become as small as one value $E$ and other
sets as small as another value $\bar E$. The methods used to prove
such bounds would be very similar to those used in this paper, but the
argument would clearly be rather more complicated. We therefore believe
the above explanation is more suitable for this paper.} But finally,
we see that
$\Lambda P \log({\Lambda_R\over
\Lambda}) \leq E P\log({\Lambda_R\over E})$ for $E\geq \Lambda$ and
this term may be written as
$\Lambda_R^{3-2m+e}E^{-e-j}P\log({\Lambda_R \over E})$.

Substituting the bounds in iii) and ii) and the results derived above
into \eIxi\ and looking at the case where $E \geq \Lambda'$, we obtain
\eqn\eIxvi{\eqalign{\Biggl \Vert{\partial \over \partial\Lambda'} \biggl(
\partial^{j}_p V^r_{2m}(\Lambda') \biggr)
\Biggr\Vert_{\Lambda_R}^{E,e}\!\!\leq
\Lambda_R^{3-2m +e}&E^{-e-j} P\log\biggl({\Lambda_R\over E}\biggr)\cr
&+ \sum \Lambda_R^{4-2m +e_A+e_B} E^{-1-e_A -e_B-j}
P\log\biggl({\Lambda_R \over E}\biggr).\cr}}

The first term comes from the first term on the right--hand side
of \eIxi, as described above.
The second term on the right--hand side of \eIxvi\ comes from those terms
on the right of \eIxi\ explicitly involving two vertices,
where the vertex on which resides the original
largest exceptional set  we call vertex $A$, letting the other
be vertex $B$, and where the sum is over
the possible values of $e_A +e_B$, and the maximum
value of $e_A +e_B$ is $e+1$. Since $e$ is odd, we see
that so too is $e_A$. However, we are not able to
simply substitute $e_A +e_B= e+1$ into \eIxvi\ and take this to be the
dominant term, because we derived the result $e_A +e_B=e+1$ by
assuming that the largest exceptional set on vertex $B$ contained only
one momentum. If this is the case then $e_B =2m_B-1$, where $2m_B$ is
the number of legs on vertex $B$. But from iii) the greatest value of
$e_B$ as far as the bounding is concerned is $2m_B-3$ (even including
the special case $m_B=1$ when we consider the result above);
all terms including two vertices on the right--hand side of
\eIxi\ contribute the same effective value of $e_A +e_B=e-1$ so
long as $e_A +e_B \geq e-1$, so the dominant term gives
$e_A + e_B =e-1$. Thus both terms on
the right--hand side of \eIxvi\ give an equally large contribution and
we have
\eqn\eIxxiii{\Biggl \Vert{\partial \over \partial\Lambda'} \biggl(
\partial^{j}_p  V^r_{2m}(\Lambda') \biggr)
\Biggr\Vert_{\Lambda_R}^{E,e}\leq
\Lambda_R^{3-2m +e}E^{-e-j} P\log\biggl({\Lambda_R \over
E}\biggr).}
It is relatively easy to convince oneself that the `other terms' in
\eIxi, which have a factor of $\Lambda_R^{-3}$ rather than $E^{-3}$, contribute
terms the same as those
explicitly displayed, but with factors of $(E/\Lambda_R)$ to a
positive power, and may therefore be absorbed into the leading
term. If we did not split the exceptional sets we would have
a maximum effective value of $e_1+e_2 =e-1$, as compared to a maximum
of $e_1+e_2=e$ if we had split them (actually, only if we split
the largest set), and thus have lost a factor of $(\Lambda_R/E)$.
However, we gain a factor of $(\Lambda_R/E)^3$, or more if we have
derivatives, from the propagator linking the two vertices.

Considering $\Lambda'> E$, as it will be for part of the range of
integration, then, as we see
from the lemma, the value of $E$ becomes
unimportant; the bound on the vertex is the same for all $E$ and is as
though $E$ were equal to
$\Lambda'$. Thus, for $\Lambda'>E$, we replace \eIxxiii\ by
\eqn\eIxx{\Biggl \Vert{\partial \over \partial\Lambda'} \biggl(
\partial^{j}_p  V^r_{2m}(\Lambda') \biggr)
\Biggr\Vert_{\Lambda_R}^{E,e}\leq
\Lambda_R^{3-2m+e}(\Lambda')^{-e-j}P\log\biggl({\Lambda_R \over
\Lambda'}\biggr),}
We can therefore evaluate \eIxix\ by splitting the integral to find
\eqn\eIxxi{\Vert
\partial^{j}_p  V^r_{2m}(\Lambda)
\Vert_{\Lambda_R}^{E,e}\leq \Lambda_R^{4-2m-j} +
\Lambda_R^{3-2m+e}E^{1-e-j}P\log({\Lambda_R \over E}),}
which verifies iii) for
$\Lambda\in [\Lambda_m,\Lambda_R]$, $e\geq \Lambda$, and $e$ odd
and $\leq 2m-3$.
If $E<\Lambda$ then we still obtain \eIxix, but in this case
$E<\Lambda'$ for the whole range of integration. Thus, we can use the
same arguments as those above in order to verify iii)
for $\Lambda \in [\Lambda_m,\Lambda_R]$, $E<\Lambda$, and $e$ odd and
$\leq 2m-3$.

The extensions to even $e$ and
$e>2m-3$ are now very easy. If $e$ is even then so is $e+2$ and $e_A$.
Thus, both of the explicitly written contributions on the right--hand
side of \eIxvi, or its equivalent for $E<\Lambda'$, will increase
their power of $E$, or $\Lambda'$, by one and, correspondingly,
decrease
their power of $\Lambda_R$ by one. It is clear from the rest of the above
argument that this increase in the power of $E$ and decrease in the
power of $\Lambda_R$ will carry through to the bound on $\Vert
\partial^{j}_p  V^r_{2m}(\Lambda)
\Vert_{\Lambda_R}^{E,e}$, and thus \eIiii\ is verified for even $e$ for
$e\leq2m-3$. If $e>2m-3$ then since $2m_A =2m$ at the most, then the
effective value of $e_A$ must decrease by
${e-2m+3}$. Again, it is clear that this will carry
through to the bound on $\Vert \partial_p^j
V^r_{2m}(\Lambda)\Vert_{\Lambda_R}^{E,e}$ and consequently the bound on the
$2m$--point vertex
remains the same for $e>2m-3$ as it was for $e=2m-3$; exactly the
result we want. (If we were to split the largest exceptional set
between the vertices, then it is easy to see that the maximum effective
value of
$e_1 +e_2$ is unchanged in this case, and such splitting still leads
to sub--dominant contributions.) We have therefore now verified iii) for all
$\Lambda \in [\Lambda_m,\Lambda_R]$ and for all $e$.

The case $\Lambda \in [0,\Lambda_m]$ is treated in a similar manner.
Considering $E\geq \Lambda_m$ and odd $e$ which is $\leq2m-3$, and integrating
from $\Lambda$ up to
$\Lambda_m$, taking norms, and using the derived
boundary condition obtained by evaluating \eIiii\ at $\Lambda_m$ we obtain
\eqn\eIxxii{ \Vert
\partial^{j}_p V^r_{2m}(\Lambda)
\Vert_{\Lambda_R}^{E,e}\leq \Lambda_R^{3-2m +e
}E^{1-e-j}P\log\biggl({\Lambda_R\over E}\biggr) +
\int_{\Lambda}^{\Lambda_m}d\Lambda' \Biggl \Vert{\partial \over
\partial\Lambda'}  \biggl(
\partial^{j}_p V^r_{2m,e}(\Lambda') \biggr)
\Biggr\Vert_{\Lambda_R,E}.}
This time, since $\Lambda'$ is always less than $E$, we bound the
integrand in the second term on the right obtaining \eIxvi\ for all $\Lambda'$.
We can then easily evaluate the integral in \eIxxii\ to find that the
second term on the right is
$(\Lambda_m-\Lambda)\Lambda_R^{2+2e}E^{1-2e-j}P\log({\Lambda_R \over E})$.
This is clearly less than or equal to the bound in iii) for
all $\Lambda$ and $E$, and substituting it into \eIxxii\ we
immediately verify iii) for $\Lambda \in [0,\Lambda_m]$, $E\geq \Lambda_m$,
$e$ odd and $\leq 2m-3$. If $E<\Lambda_m$ then the argument is exactly the
same once we replace $E$ by $\Lambda_m$. Thus, we easily verify iii)
for $\Lambda \in [0,\Lambda_m]$,
$E<\Lambda_m$, $e$ odd and $\leq 2m-3$.
For both $E\geq \Lambda_m$ and $E<\Lambda_m$ the extension to even $e$, or
$e>2m-3$ is obvious.

It is now comparatively simple to verify ii) for the irrelevant
vertices. Considering any $\Lambda \in [0,\Lambda_R]$,
integrating from $\Lambda$ up to $\Lambda_R$, taking norms, and using the
derived boundary condition obtained by evaluating i) at $\Lambda_R$ we obtain
\eqn\eIxxiv{ \Vert
\partial^{j}_p V^r_{2m}(\Lambda)
\Vert_{\Lambda_R}\leq \Lambda_R^{4-2m-j} +
\int_{\Lambda}^{\Lambda_R}d\Lambda' \Biggl \Vert{\partial \over
\partial\Lambda'}  \biggl(
\partial^{j}_p  V^r_{2m}(\Lambda') \biggr)
\Biggr\Vert_{\Lambda_R}.}
Again, we are able to bound the integrand in the second term on the right using
the bounds we already have on vertices at
orders $r-1$ and less in $g$ and vertices of $m+1$ or more legs, i.e.
those in ii) and iii) (the bound iv) not being needed this time).
So substituting the bounds in ii) and iii) into \eIxii\ and using our
result for the first term on the right--hand side we see that the
integrand in \eIxxiv\ is $\leq \Lambda_R^{3-2m-j}$ for all
$\Lambda' \in [0,\Lambda_R]$.
Substituting this result into \eIxxiv\ and performing the integral it
is very easy to see that \eIii\ is verified.
Thus, we have verified lemma 5 for all irrelevant vertices at order $r$
in $g$.

c) The proof of the lemma for the relevant vertices
proceeds in a similar way to the proof of lemma 1 for the
relevant vertices in \S 2.2 in \REFT. We first consider the 4--point
vertex. Since the momenta at which the renormalization condition for
this vertex was set are non--exceptional, we can use \eIxii\ and the
bounds already obtained for the vertices at lower order in $g$ or
equal order in $g$, but with greater $m$, to write
\eqn\eIxxix{\Bigl\vert {\partial \over \partial \Lambda}
V^r_4(P_i;\Lambda) \Bigr\vert \leq  \Lambda_R^{-1},}
and subsequently,
\eqn\eIxxx{\vert V^r_4(P_i;\Lambda_R) -
V^r_4(P_i;0) \vert
 \leq \int^{\Lambda_R}_{0} d
\Lambda' \Biggl\vert {\partial \over \partial \Lambda'}
V^r_4(P_i;\Lambda')\Biggr\vert\leq c.}
So, using the renormalization condition on $V^r_4(P_i;0)$, we
can say that $\vert V^r_4(P_i;\Lambda_R)\vert \leq c$
and we have a bound on the vertex
defined at $\Lambda_R$ for the particular
momenta at which the renormalization condition is set. Using Taylor's
formula, as in \REFT, for $\Lambda =\Lambda_R$ we can verify
ii) and iii) (which are identical for $\Lambda=\Lambda_R$) and
thus obtain a boundary condition on
the vertex at $\Lambda_R$; $\Vert V^r_4(\Lambda_R)\Vert\leq c$.
It is now straightforward to verify ii) and iii) for
$V^r_4(\Lambda)$ for $\Lambda \in [0,\Lambda_R]$ in exactly the
same way as these bounds were verified for the irrelevant vertices.
Thus, ii) and iii) are verified for the four--point vertex at order
$r$ in $g$.

The verification of ii) and iii) for the second momentum derivative of
the two--point vertex is performed in exactly the same manner as for
the four--point vertex. The cases of the two--point vertex and its
first momentum derivative are not quite so simple due to the fact that
they obey a different lemma to all the other vertices if
$\Lambda\leq\Lambda_R$ and $p\leq\Lambda_R$. However, iv) may be
verified using Taylor's formula in a rather simple manner.

Using the result
$\Vert \partial^{2}_p V^r_{2}(\Lambda)\Vert_{\Lambda_R}^{E,1}\leq
P\log({\Lambda_R\over \bar \Lambda})$, i.e.
iii) for the second derivative of the two point function, we
immediately see that for all $p \leq \Lambda_R$ that $\partial^{2}_{p}
V^r_{2}(p,-p;\Lambda)\leq P\log({\Lambda_R\over E})$ for $p =E\geq
\Lambda$ and that $\partial^{2}_{p}
V^r_{2}(p,-p;\Lambda)\leq P\log({\Lambda_R\over \Lambda})$ for $p
\leq \Lambda$. We may construct
$\partial_{p_{\mu}}V^r_2(p,-p;\Lambda)$ using Taylor's formula;
\eqn\eIxxxii{ \partial_{p_{\mu}}V^r_2(p,-p;\Lambda)=p_{\nu}
\int_0^1 d\rho
\partial_{k_{\nu}}\partial_{k_{\mu}}V^r_2(k,-k;\Lambda),}
where $k=\rho p$. Taking the modulus of both sides and adopting the
same notation as in Appendix B of \REFT, i.e. letting $\bar p$ denote any
particular component of momenta and noting that the sum over
components does not change our qualitative result, we obtain
\eqn\eIxxxiii{ \vert\partial_{p_{\mu}}V^r_2(p,-p;\Lambda)\vert \leq
\vert \bar p \vert
\int_0^1 d\rho
\vert\partial_{\bk}\,\partial_{k_{\mu}}V^r_2(k,-k;\Lambda)\vert.}
If $E\leq \Lambda$, the right--hand side of this inequality is
clearly $\Lambda P\log({\Lambda_R\over\Lambda})$. If $\Lambda \leq
E\leq \Lambda_R$ then the right--hand side is $\leq E \int_0^1 d \rho P\log
({\Lambda_R\over \rho E})$. The integral over $\rho$ gives
$P\log({\Lambda_R\over
E})$ and the right--hand side of the inequality is
$E\,P\log({\Lambda_R\over E})$. So, for  all $0\leq E \leq \Lambda_R$ we
have verified iv) for the first momentum derivative of the
two--point function at order $r$ in $g$ simply by using Taylor's formula
(Appendix B of \REFT\ not
being required). This is clearly true for $\Lambda \in [0,\Lambda_m]$
as well as for $\Lambda \in [\Lambda_m,\Lambda_R]$.

In order to verify ii) for $\partial_p V^r_2(\Lambda)$ we must again use
Taylor's formula in conjunction with the techniques of Appendix B. In
fact, since $\partial_p V^r_2(\Lambda)$ vanishes at zero momentum simply
due to Lorentz invariance, we can construct it entirely from
$\partial^2_p V^r_2(\Lambda)$ using Taylors formula about zero momentum.
Doing this for $\Lambda=\Lambda_R$ we verify ii) for $\partial_p
V^r_2(\Lambda_R)$. It is then easy to verify ii) for all $\Lambda \in
[0,\Lambda_R]$ in the same way as for $V^r_4(\Lambda)$ and the
irrelevant vertices.

Finally we consider the two-point vertex with no momentum derivatives.
We look at $\Lambda \in [0,\Lambda_R]$ for the vertex
at the momentum where the renormalization condition is set.
Integrating from $\Lambda$ down to $0$, we obtain
\eqn\eIxxxiv{\vert V^r_2(\tilde P,-\tilde P;\Lambda) \vert
\leq \vert V^r_2(\tilde{P},-\tilde{P};0) \vert
 \leq \int^{\Lambda}_{0} d
\Lambda' \Biggl\vert {\partial \over \partial \Lambda'}
V^r_2(\tilde{P},-\tilde{P};\Lambda')\Biggr\vert.}
Since this momentum is exceptional, we must use \eIv, where in this
case $E\sim\Lambda_m$, and the bounds already obtained for the two
point vertex at lower order in $g$ and the four--point vertex at
equal order in $g$. The first term on the right of \eIv\ is then
clearly $\leq \Lambda P\log({\Lambda_R \over \Lambda})$. The maximum value
of $\Lambda^{-3}K'_{\Lambda}(p_1)$ is $\leq \Lambda^{-3}$, and from iv)
the maximum values
of $V^s_2(\Lambda)$ and $V^{r-s}_2(\Lambda)$ are both $\leq
\Lambda^2P\log({\Lambda_R \over \Lambda})$. The second term on the
right--hand side of \eIv\ is therefore also
$\leq\Lambda P\log({\Lambda_R/\Lambda})$ (or
$\Lambda_mP\log(\Lambda_R/\Lambda_m)$ when $\Lambda\leq\Lambda_m$).
Substituting this into
\eIxxxiv, performing the integral and using the renormalization
condition \eIxv\ on $V^r_2(\tilde{P},-\tilde{P};0)$, we find
that $\vert V^r_2(\tilde{P},-\tilde{P};\Lambda)\vert \leq
\Lambda^2 P\log({\Lambda_R\over \Lambda})$, and we have a bound on the vertex
defined at $\Lambda$ for the particular
momenta at which the renormalization condition is set.
We can now verify iv) using the Taylor formula, as for the first
derivative of the two--point function, i.e. using the equation
\eqn\eIxxxv{ \vert V^r_2(p,-p;\Lambda)\vert \leq
\Lambda^2P\log\biggl({\Lambda_R \over \Lambda}\biggr) + \vert
\bar p + \bar q\vert
\int_0^1 d\rho
\vert\partial_{\bk}V^r_2(k,-k;\Lambda)\vert,}
with $k = q + \rho (p-q)$. The argument is clearly the
same as that used for the first momentum derivative if
$q=\tilde P=0$, with the additional term from
the value of the vertex at $p=\tilde{P}$ obviously satisfying
\eIxiix. We can see that the argument is
essentially unchanged if $\bar p \not= 0$, since it is only of the
order of $\Lambda_m$. Thus, $\Lambda$ or $E$ in \eIxxxv\ will only be replaced
with $\Lambda \pm \Lambda_m$ or $E \pm \Lambda_m$ respectively, in comparison
with the $\tilde{P}=0$ case, and these are of order $\Lambda$ and
$E$, and act the same way as far as bounds are
concerned. iv) is therefore verified for the two--point vertex at
order $r$ in $g$ for $\Lambda \in[0,\Lambda_R]$.

We may verify ii) for the two--point vertex in a straightforward
manner. Since $E P\log({\Lambda_R\over E}) \leq \Lambda_R$ when $E\leq
\Lambda_R$ we know that
$\Vert\partial_{p} V^r_{2}(p,-p;\Lambda)\Vert_{\Lambda_R}^{E,1}\leq
\Lambda_R$
for all $p_1$. We can therefore use Taylor's formula and the result in
appendix B of \REFT\ to prove that $\Vert
V^r_{2}(\Lambda)\Vert_{\Lambda_R}^{E,1}\leq \Lambda^2_R$. This
verifies \eIii\ for the two--point vertex, and is consistent with, but
weaker than, \eIxiix\ for this vertex. Thus, we have now verified
ii), iii) and iv) for all the vertices at order $r$ in $g$.

\medskip

d)  Using the
derived boundary condition for the relevant couplings at
$\Lambda_R$ for some momenta with magnitude $\leq \Lambda_R$, e.g. all
momenta zero or momenta equal to those at which the renormalization
conditions at $\Lambda=0$ were set, we
can verify i) for the relevant vertices by integrating the
coupling constants from
$\Lambda\in[\Lambda_R,\Lambda_0]$ down to $\Lambda_R$ and using Taylor's
formula, as in \S 3.1; working downwards in $m$ and, for given $m$,
downwards in number of derivatives.  Thus, lemma 5 is verified for the relevant
vertices at order
$r$ in $g$, and therefore for all vertices at this order. By
induction, since it is trivially satisfied at zeroth order in $g$,
the lemma is therefore true at all orders in $g$. \blackbox

\medskip

In particular, we consider ii), iii) and iv) at
$\Lambda=0$. Remembering the direct relationship \evsc\ between the vertices
defined at $\Lambda=0$ and the Green's functions we find that for $e=0$
\eqn\eIxxxvi{ \bigl\Vert \partial^j_p \tilde
G^{c}_{2m}(p_1,\ldots,p_{2m};\Lambda_0,\lambda_i) \bigr\Vert_{\Lambda_R} \leq
\Lambda_R^{4-2m-j}.}
For $e\geq 1$,  and $m>1$ or $m=1$, $j>1$,\foot{$\bar \Lambda$ is now
effectively $\max(E,\Lambda_m)$ since $\Lambda=0$.}
\eqn\eIxxxvii{\bigl\Vert \partial^{j}_p
\tilde G^{c}_{2m}(p_1,\ldots,p_{2m};\Lambda_0,\lambda_i)
\bigr\Vert_{\Lambda_R}^{E,e}\leq\cases{
 \Lambda_R^{4-2m-j}
 \Bigl({\Lambda_R\over \bar\Lambda}\Bigr)^{e+j-1}
 P\log \Bigl({\Lambda_R \over \bar\Lambda}\Bigr) & $e$ odd,\cr
 \Lambda_R^{4-2m-j}
 \Bigl({\Lambda_R\over \bar\Lambda}\Bigr)^{e+j-2}
 P\log \Bigl({\Lambda_R \over \bar\Lambda}\Bigr) & $e$ even.\cr}}
up to a maximum of $e=2m-3$, where for $e$ greater than this value (even for
$e=1$, $m=1$) the bound
is $\bar\Lambda^{4-2m-j} P\log ({\Lambda_R \over \bar \Lambda})$.
For $e=1$, $m=1$ and $j<2$, we have
\eqn\eIxxxiix{ \max_{p \leq E}\bigl\vert \partial^j_p \tilde
G^{c}_{2}(p,-p;\Lambda_0,\lambda_i)
\bigr\vert \leq
\bar\Lambda^{2-j} P\log \biggl({\Lambda_R\over \bar\Lambda}\biggr).}

In fact we can actually improve the right--hand side of these bounds if we
change the way in which we take the norms slightly, and consider the
momentum derivatives more carefully. All the norms considered above
were entirely general; we never considered which of the
momenta became exceptional, only that certain types of sets did.
For the vertices themselves this is all it makes sense to do anyway since
the vertices are invariant under permutations of momenta due to the
Bose symmetry. However, when we differentiate with respect to a
particular momentum we break this Bose symmetry, and should thus use a
norm which distinguishes between different momenta, taking account of
whether the derivatives are with respect to those momenta that are
within exceptional sets, or not. It is then possible to show that the
bounds \eIxxxvii\ may be improved by a factor of
$\bar\Lambda/\Lambda_R$ for each momentum derivative which acts on a
momentum which is not in some irreducible exceptional set.

We could also derive very similar bounds on a theory with
no $Z_2$ symmetry. The bounds for all vertices with no exceptional
momenta, or with numbers of legs
plus derivatives, $n+j$, greater than 3, would be as in equations \eIxxxvi\ and
\eIxxxvii, with $2m+j$ obviously replaced by $n+j$. We would again
have \eIxxxiix\ for the two--point function, but would also need the bound
\eqn\eIxxxiixz{ \max_{p_1, p_2 \leq E}\bigl\vert \tilde
G^{c}_{3}(p_1,p_2,-p_1-p_2;\Lambda_0,\lambda_i)
\bigr\vert \leq
\bar\Lambda P\log \biggl({\Lambda_R\over \bar\Lambda}\biggr),}
which can only be obtained by setting the renormalization condition
that the three point vertex has magnitude $\sim \Lambda_m$ for
external momenta with magnitude $\sim \Lambda_m$. In particular, as $m
\to 0$ the three point function must be set to be zero for zero
external momenta.\foot{For a conventional quantum field theory this
would mean that the massless theory must be $Z_2$--symmetric, since
all the low energy renormalization conditions maintain this symmetry. In the
effective theory we may still have $Z_2$ breaking terms in the form of
irrelevant bare vertices. Hence, due to the insensitivity of the low
energy theory to these irrelevant vertices the $Z_2$ symmetry is only
very weakly broken at low energies.}
This is a reflection of the well--known result that infrared finite
massless
theories are not allowed to have super--renormalizable couplings.
However, we see this more clearly as a restriction
on the three--point Green's function than one on the
bare three--point vertex.

\subsec{Infrared Finiteness.}

In \S 2.4 we have provided a rather detailed description of
the infrared behaviour in the Euclidean region of
the theory defined in \S 2.1, where `infrared' is taken to mean
momenta with magnitudes below those at
which the renormalization conditions (other than that on the mass)
were set, i.e. below $\Lambda_R$. We can now consider the limit
$m\rightarrow 0$ in order to investigate the behaviour of a strictly massless
theory. We see immediately that all amputated connected Green's
functions at non-exceptional momenta
remain finite in this limit to all orders in perturbation theory,
and we can also see how quickly the Green's
functions may diverge for different combinations of momenta tending
to zero.

The proof of convergence in the low mass case is a straightforward
combination of the boundedness argument in the previous section and
that in \S 2.3 of \REFT. The
$\Lambda_0$--derivative of the Green's functions away from exceptional
momenta is infrared finite at low energies, and is weighted by an
extra factor of
$({\Lambda_R\over\Lambda_0})P\log({\Lambda_R/\Lambda_0})$;
as exceptional momenta are approached, the $\Lambda_0$--derivative of
the Green's functions displays the same degree of infrared divergence
as the Green's function itself. Similarly if we were to introduce
natural bare irrelevant couplings (as in \S 3.1 of \REFT), the nature
of the infrared
divergences is unchanged; the massless theory is convergent and
universal in just the same way as the massive theory. It can also be
systematically improved (see \S 3.2 of \REFT) by setting more low
energy renormalization conditions (obviously in a manner consistent
with the bounds, and for momenta with magnitudes $\sim \Lambda_R$);
the low energy Green's functions will then only change by amounts of
order the Green's functions themselves, weighted by further
powers of $({\Lambda_R/\Lambda_0})$.
Thus, we can draw exactly the same conclusions for a massless theory as we
did for the massive theory; if we set renormalization conditions on
the low energy relevant couplings which are independent of the
naturalness scale, then the low energy physics is extremely insensitive
to both the value of this naturalness scale and to the irrelevant
bare couplings, provided only that the naturalness scale is much
larger than the scale of the physics.

Since the analytic structure of our effective theory allows a
well--defined analytic continuation (as discussed in \S 2.4 of \REFT)
we may also consider the significance of our results for Minkowski space
Green's functions. Using the Landau rules one may readily demonstrate that
singularities due to the vanishing of masses only occur due to the
momenta associated with internal lines becoming either soft or
collinear \rIRii. (For an accessible discussion of this result see
\ref\rster{G.~Sterman,
``An Introduction To Quantum Field Theory'', Cambridge University
Press, Cambridge, 1993.}.)
However, this simple result does not distinguish between
internal loop momenta and external momenta, and thus does not
tell us whether or not diagrams may in fact be divergent due to internal
momenta only becoming soft and/or collinear (indeed in two dimensions
we know by experience that they do). In this context, our result proves that in
four dimensions Green's functions may not become divergent in the massless
limit simply due to internal lines becoming soft and/or collinear,
and therefore that any singularities must result from the
the external momenta becoming soft and/or collinear.
The inequality \eIxxxvii\ then shows us  the type of divergences we
may (and do) obtain when partial sums of external
momenta become soft.

{}From the Landau rules we also expect (and indeed obtain)
singularities on the boundary of the physical region when external
momenta become collinear. However, by avoiding these singularities it
is possible to perform the same type of analytic continuation as
described in \S 2.4 of \REFT. We may thus see that
all Green's functions away from the physical singularities are
strictly finite, and therefore so are their discontinuities across the
cuts (which are traditionally taken to lie along the timelike
axis).\foot{This result for the photon two point function, which is
true for all timelike momenta, may be used, for example,
to show that the total decay rate for a virtual photon (obtained
for example from $e^{+}e^{-}$-annihilation) into quarks plus
arbitrary numbers of soft gluons is finite.} Also this
analytic continuation shows that the Green's functions on
the boundary of the physical region are only very weakly dependent on
$\Lambda_0$ and the renormalization conditions on irrelevant
operators, for energies much less than $\Lambda_0$.
For example, if we were to perform an analytic continuation
from a Euclidean region
where all partial sums of momenta have magnitudes $\sim \Lambda_R$ and
analytically continue to the boundary of the physical region along a
path which avoids the singularities given by the Landau rules by a
distance $\sim \Lambda_R$, we can
also see that $\partial^j_p \tilde
G^{c,r}_{2m}(p_1,\ldots,p_{2m};\Lambda_0,\lambda_i) \leq
\Lambda_R^{4-2m-j}$ for partial sums of momenta with magnitudes of
order $\Lambda_R$ on the boundary of the physical region.

Using the analytic continuation we can also consider the S-matrix.
At first sight it appears that this is not strictly well--defined,
since there are singularities for processes involving
either very soft particles or energetic but collinear particles.
Nevertheless, the rather formal argument of \ref\rLN{T.~D.~Lee and
M.~Nauenberg \PR \vyp{B133}{1964}{1549}.} proves that for any theory
which is unitary for finite mass (as our effective theory is, as
was shown in \S 5.3 of \REFT) the
transition probabilities, and hence cross--sections, for particle
scattering are always finite in the massless
limit provided that one sums over degenerate initial and final
states containing arbitrary numbers of particles.\foot{A rather less
formal proof of this theorem, using cut diagrams, is presented in
\rster. Since this proof proceeds from the cutting relations, we may
adapt it to our effective theory by employing the `unitary
representation' discussed in \S 5.3 of \REFT, in which the propagator
assumes its usual unregulated form. Then when the cut lines go
on--shell, the regulating functions at the ends of these lines are set
to unity, and thus their essential singularity presents no obstruction
to the proof.} Our bounds on the Green's functions may thus also be
applied to S--matrix elements.

\newsec{Ultraviolet Behaviour.}

\subsec{Weinberg's Theorem.}

Within the conventional formulation of quantum field theory Weinberg's
theorem\rwt\ is a result telling us about the behaviour of Euclidean
Green's functions when all the external momenta have magnitudes
much greater than the mass of the
particles (the `deep Euclidean' region).  It states that
if the magnitude of all partial sums of the
momenta of a Green's function are of the order $E$, then a given
amputated Green's function is bounded by a constant times $E^{d}$ up to
logarithmic corrections, where $d$ is the power given by naive dimensional
analysis ($d=4-2m$ for our example of the scalar field theory with
$Z_2$ symmetry.) In our case we have a naturalness scale $\Lambda_0$,
which we take to be finite, and thus we can only hope to show
that Weinberg's theorem is true for energy scales up to
$\Lambda_0$.

In a certain sense, we have already proved this result in the previous
section. Setting $\Lambda_R=E$ in \eIxxxvii\ we find that the
value of the Green's functions (and their momentum derivatives) are
bounded by $E$ to the appropriate power, without even the
logarithmic corrections of $P\log(E/\Lambda_m)$ we might naively
expect. We made no restrictions on the value of the scale
$\Lambda_R$, other than it be greater than $m$, so it can
span the full range of values for which we would expect Weinberg's
theorem to hold. However, since we were interested in
the behaviour of the theory in the massless limit we chose to
specify that the renormalization
conditions on the second momentum derivative of the two--point
function and on the four--point function were set for momenta (and
partial sums of momenta in the latter case) with magnitudes of order
$\Lambda_R\gg m$. Here, on the other hand, we are interested not in
the massless limit, but in the behaviour of Green's functions for a
massive theory in the deep Euclidean region. We thus
want to set renormalization conditions for momenta not of order
$E$, but rather of order $m$. In particular, we want to be able to set
on--shell renormalization conditions, as described in \S 2.4 of \REFT.
The arguments of \S 2 must then be altered in order to prove
Weinberg's theorem.

\medskip

We define our theory in the same way as in \S 2.1 of \REFT, i.e. in
exactly the same way as in \S 2.1 above except for the
renormalization conditions \eIxv\ on
$[\partial_{p_{\mu}}\partial_{p_{\nu}}V^r_2(p,-p;0)
\vert_{p=P_0}]_{\delta{\mu\nu}}$
and $V^r_4(P_1,P_2,P_3,P_4;0)$, which are now set with the
$P_i$ having all having magnitudes of order $\sim m$. We also define
a scale $\Lambda_H$,
which is the scale at which we wish to investigate the physics ($\Lambda_R$ is
the scale at which the renormalization conditions are set, so in
this section $\Lambda_R\equiv\Lambda_m$).
Exceptional momenta are then defined in exactly the same way as in \S 2.2
(once we replace $\Lambda_R$ by $\Lambda_H$), as are the
restricted norms. This means that
the bounded flow equation for $\Lambda \in [\Lambda_H,\Lambda_0]$ is
the same as \exxxxvii\ in \S 1, while the bounded flow equations for
$\Lambda\in[\Lambda_m,\Lambda_H]$ and $\Lambda \in [0,\Lambda_m]$ are
the same as \eIxi, \eIv\ and \eIxii\ in \S 2.3. If we were to suggest
that our
lemma for the bounds on the vertices was identical to lemma 5
however, we would find that the proof would break down as soon as we
tried to prove the lemma for the relevant vertices (the first
obstruction would be that \eIxxix\ would be no longer true if the $P_i$
had magnitudes of order $\Lambda_m$). We must thus set up a new
lemma, fully expecting there to be logarithmic corrections to the
bounds, even when $E=\Lambda_H$: we thus propose
\medskip
\noindent{\it Lemma 6:}

\noindent i) For all $\Lambda \in [\Lambda_H,\Lambda_0]$,
\eqn\eWi{ \bigl\Vert \partial^j_p V^r_{2m}(\Lambda) \bigr\Vert_{\Lambda} \leq
\Lambda^{4-2m-j} \Biggl( P\log \biggl( {\Lambda \over \Lambda_m} \biggr)
+ {\Lambda \over\Lambda_0} P\log \biggl({\Lambda_0 \over
\Lambda_m}\biggr)\Biggr).}
\noindent ii) For all $\Lambda \in [0,\Lambda_H]$, and for
$e=0$, we have
\eqn\eWii{ \bigl\Vert \partial^j_p V^r_{2m}(\Lambda) \bigr\Vert_{\Lambda_H}
\leq
\Lambda_H^{4-2m-j}P\log\biggl({\Lambda_H \over \Lambda_m}\biggr).}
\noindent iii) For all $\Lambda \in [0,\Lambda_H]$ and for
$1\leq e\leq 2m-3$, $m> 1$, or $m=1$, $j>1$
\eqn\eWiii{ \bigl\Vert \partial^{j}_p  V^r_{2m}(\Lambda)
\bigr\Vert_{\Lambda_H}^{E,e} \leq\cases {
 \Lambda_H^{4-2m-j}
 \Bigl({\Lambda_H\over \bar\Lambda}\Bigr)^{e+j-1}
 P\log \Bigl({\Lambda_H \over \Lambda_m}\Bigr) & $e$ odd,\cr
 \Lambda_R^{4-2m-j}
 \Bigl({\Lambda_H\over \bar\Lambda}\Bigr)^{e+j-2}
 P\log \Bigl({\Lambda_H \over \Lambda_m}\Bigr) & $e$ even,\cr}}
where $\bar\Lambda={\rm max}(E,\Lambda,\Lambda_m)$.
For $e>2m-3$ the bound is simply
$\bar\Lambda^{4-2m-j} P\log ({\Lambda_H \over \Lambda_m})$.\foot{
We notice that the
bound iii) for the case when $e=2m-1$, and thus when all momenta may become
small individually is greater than the bound obtained in \REFT
(lemma 1) for all momenta being of order $m$, i.e. $\Vert \partial^j_p
V^r_{2m}(\Lambda)\vert \leq
m^{4-2m-j}$ for all $p_i\sim m$, $\Lambda\leq m$. This is
because the bound in lemma 6 is for the case when all momenta
`may', but need not, become small, so the bounds cannot decrease
as $e$ increases. The fact that we have the bound of
$\Lambda_m^{4-2m-j}P\log({\Lambda_H\over \Lambda_m})$ for $e=2m-3$
means then that
the bound for $e=2m-1$ must be at least as large. After proving the bounds in
lemma 6 it is possible to go back and investigate the case where all
momenta actually are of order $m$ with an equation similar to \eIxi, and verify
the expected result. (The bound over all momenta being needed first because of
the one
term on the right--hand side with an integral over all momenta.) We
leave this verification to the more dedicated reader.}

\noindent iv) The case $\Lambda \in [0,\Lambda_H]$ and  $e= 1$,
$m=1$, $j=0,1$ is a little different, as in lemma 5. Again,
we do not have a restricted norm, but an inequality similar
to \eIiv,
\eqn\eWv{ \max_{p \leq E}\bigl\vert \partial^j_p V^r_{2}(\Lambda)
\bigr\vert \leq
\bar\Lambda^{2-j} P\log \biggl({\Lambda_H\over \Lambda_m}\biggr).}

\medskip

Using equations \eIxi, \eIv\ and \eIxii\ we may set about proving lemma 6.
As for all previous lemmas the proof is by induction, and the
induction scheme is the same as for lemma 1. We assume that the lemma
is true up to order $r-1$ in the expansion coefficient $g$, and that
at first order in $g$ it is true down to $m+1$.

\medskip

a) We first consider the irrelevant vertices for $\Lambda\in
[\Lambda_H,\Lambda_0]$. Just as in \S 2.4a, the flow equation and the boundary
conditions on the irrelevant vertices are exactly the same as they
were for lemma 1 in \REFT. The bound is also the same except for
the fact that $\Lambda_R$ is replaced by $\Lambda_m$. It is easy to
see that this does not change the argument used in the proof of
lemma 1 as long as $\Lambda_H\geq\Lambda_m$. Thus, we may
verify i) for $\Lambda$ in this range simply by
using the flow equation and integrating from $\Lambda$ up to $\Lambda_0$.

\medskip

b) The step for the irrelevant vertices with $\Lambda \in
[\Lambda_m,\Lambda_H]$ and $\Lambda \in [0,\Lambda_m]$ is very similar
to that in \S 2.4b.  We
first consider iii) for $\Lambda \in [\Lambda,\Lambda_H]$ and $E\geq \Lambda$,
with $e$ odd and $\leq 2m-3$.
Integrating from $\Lambda$ up to $\Lambda_H$, taking norms, and using the
derived
boundary condition obtained by evaluating \eWi\ at $\Lambda_H$ we obtain
\eqn\eWxi{ \Vert
\partial^{j}_p V^r_{2m}(\Lambda)
\Vert_{\Lambda_H}^{E,e}\leq \Lambda_H^{4-2m-j}
P\log\biggl({\Lambda_H\over\Lambda_m}\biggl)+
\int_{\Lambda}^{\Lambda_H}d\Lambda' \Biggl \Vert{\partial \over
\partial\Lambda'}  \biggl(
\partial^{j}_p  V^r_{2m}(\Lambda') \biggr)
\Biggr\Vert_{\Lambda_H}^{E,e}.}
We are now able to bound the integrand in the second term on the right
using the bounds in ii), iii) and iv). Using iv) we can now simplify
the last set of terms in \eIxi\ in a similar way as we did in \S 2.4b.
We thus find that
\eqn\ewVi{\max_{p_i\geq E} (\Lambda^{-3}
\vert\partial^{j_{1}}K_{\Lambda}(p_i)\vert\cdot\vert \partial^{j_{2}}_p
V^{s}_{2}(p_i;\Lambda)\vert) \leq
\bar\Lambda^{-1-j_1-j_{2}}P\log\biggl({\Lambda_H\over \Lambda_m}\biggr).}
So, once again the term
on the right--hand side of \eIxi\ containing the two--point function and its
first derivative with exceptional momentum effectively behaves like the other
explicitly written terms with two vertices.

So, substituting the bounds in iii) and ii) and the result
derived above into \eIxi\ both when $E \geq \Lambda'$ and $E\Lambda'$,
and realizing that the power
counting for $E$ works in exactly the same way as previously, we obtain
\eqn\eWxivi{\eqalign{ \Vert
\partial^{j}_p  V^r_{2m}(\Lambda)
\Vert_{\Lambda_H}^{E,e}\leq \Lambda_H^{4-2m-j} P\log\biggl({\Lambda_H\over
\Lambda_m}\biggr)+
\Lambda_H^{3-2m+e}\biggl(\int_{\Lambda}^{E} &\!d\Lambda'
E^{-e-j}P\log\biggl({\Lambda_H \over
\Lambda_m}\biggr)\cr
&\!\!\!+  \int_{E}^{\Lambda_H}\!d\Lambda'
(\Lambda')^{-e-j}P\log\biggl({\Lambda_H \over
\Lambda_m}\biggr)\biggr).\cr}}
Again, the other terms are sub--dominant.
Both integrals are trivial to evaluate, and we immediately verify \eWiii\ for
$\Lambda\in [\Lambda_m,\Lambda_H]$ and $e$ odd and $\leq 2m-3$. The
extension to $E<\Lambda$ and then to even $e$ and $e>2m-3$ is exactly
as in \S 2.4b. The case $\Lambda \in [0,\Lambda_m]$ is also treated in
an exactly analogous manner.

It is then very easy to verify ii) for the irrelevant
vertices, using the same method as was used for the $e=0$ irrelevant
vertices in \S 2.4b. In fact, it is even more straightforward than this
argument because we do not have to eliminate the logarithm. We have
therefore verified lemma 6 for all irrelevant vertices at order $r$ in
$g$.

\medskip

c) The proof of the lemma for the relevant vertices is again very
similar to that in \S 2.4c, and is, if anything, easier.
Again, we first consider the 4--point
vertex. This time the momenta at which the renormalization condition for
this vertex was set are exceptional as low as $\Lambda\sim\Lambda_m$.
We therefore use \eIxi\ and the
bounds already obtained for the vertices at lower order in $g$ or
equal order in $g$, but with greater $m$, to write
\eqn\eWxvi{\Biggl\vert {\partial \over \partial \Lambda}
V^r_4(P_i;\Lambda) \Biggr\vert \leq
\Lambda^{-1}P\log\biggl({\Lambda_H\over \Lambda_m}\biggr),}
for $\Lambda\in[\Lambda_m,\Lambda_H]$, and the same if we replace
$\Lambda^{-1}$ by $\Lambda_m^{-1}$ for $\Lambda\in[0,\Lambda_m]$. But,
\eqn\eWxviia{\vert V^r_4(P_i;\Lambda_H) \vert
\leq \vert V^r_4(P_i;0) \vert
+ \int^{\Lambda_H}_{0} d
\Lambda' \Biggl\vert {\partial \over \partial
\Lambda'}V^r_4(P_i;\Lambda')\Biggr\vert.}
Using the renormalization condition on $V^r_4(P_i;0)$, and
splitting the integral into two, one from $\Lambda=0$ to $\Lambda_m$
and the other from $\Lambda_m$ to $\Lambda_H$, we easily obtain
$ \vert V^r_4(P_i;\Lambda_H)\vert
\leq P\log({\Lambda_H\over \Lambda_m})$,
and thus have a bound on the vertex
defined at $\Lambda_H$ for the particular
momenta at which the renormalization condition is set. Using Taylor's
formula for $\Lambda =\Lambda_H$ we can verify
\eWii\ and \eWiii\ for $\Lambda =\Lambda_H$, and
 obtain a boundary condition on the vertex at $\Lambda_H$;
$\Vert V^r_4(\Lambda_H)\Vert\leq P\log({\Lambda_H\over \Lambda_m})$.
It is then straightforward to verify
ii) and iii) for $V^r_4(\Lambda)$ for $\Lambda \in
[\Lambda_m,\Lambda_H]$ and $\Lambda \in [0,\Lambda_m]$ in exactly the
same way as these bounds were verified for the irrelevant vertices.
Thus, ii) and iii) are verified for the four--point vertex at order
$r$ in $g$.

The verification of ii) and iii) for the second momentum derivative of
the two--point vertex is performed in exactly the same manner as for
the four-point vertex. As in \S 2.4c, the cases of the
two--point vertex and its
first momentum derivative are not quite so simple because they obey a
different type of bound to all the other vertices if $\Lambda\leq
\Lambda_H$ and $p\leq\Lambda_H$. However, these bounds may be verified
in a similar, but more straightforward, way as that which we used to verify
iv) in \S 2.4.

Using the result iii) we
 see that for all $p\leq \Lambda_H$ that $\vert \partial^{2}_{p}
V^r_{2}(p,-p;\Lambda)\vert\leq P\log({\Lambda_H\over \Lambda_m})$.
We may construct
$\partial_{p_{\mu}}V^r_2(p,-p;\Lambda)$ for $\Lambda\in[0,\Lambda_H]$ using
Taylor's formula.
Taking the modulus of both sides and again adopting
the notation in Appendix B of \REFT\ and letting $k=\rho p$, we obtain
\eqn\eWxix{ \vert\partial_{p_{\mu}}V^r_2(p,-p;\Lambda)\vert \leq
\vert \bar p \vert
\int_0^1 d\rho
\vert\partial_{\bk}\,\partial_{k_\mu}V^r_2(k,-k;\Lambda)\vert.}
If $p=E$, the right--hand side of this inequality is
clearly $E P\log({\Lambda_H\over\Lambda_m})$, and we have verified iv).
In order to verify ii) for $\Vert \partial_p
V^r_2(p_1;\Lambda)\Vert_{\Lambda_H}$ we can use
the result that $\Vert \partial^2_j V^r_2(p,-p;\Lambda)\Vert_{\Lambda_H}^{E,1}
\leq P\log({\Lambda_H\over \Lambda_m})$ for $E$ as low as zero, in
conjunction with Taylor's formula and the result in appendix B of
\REFT\ to show
that $\Vert \partial_p V^r_2(p,-p;\Lambda)\Vert_{\Lambda_H}^{E,1}\leq
\Lambda_HP\log({\Lambda_H\over \Lambda_m})$ for $E$ as
low as zero. This clearly verifies ii) and is consistent with iv).
Thus, ii) is verified for
$\partial_{p}V^r_{2m}(\Lambda)$ at order $r$ in $g$.

Finally we consider the two-point vertex with no momentum derivatives.
The proof of iv) for the two--point vertex is practically identical
to that of iv) in lemma 5. Again we use \eIv\
instead of \eIxi, and the argument is exactly the same.
iv) is therefore verified for the two--point vertex at
order $r$ in $g$ for $\Lambda \in[0,\Lambda_H]$.
The verification of ii) for the
two--point vertex is, again, the same as the corresponding
verification in the last section, and ii), iii) and iv)
are verified for all vertices at order $r$ in $g$.

\medskip

d)  Using the
derived boundary conditions for the relevant couplings at
$\Lambda_H$ for momenta with magnitude $\leq \Lambda_H$, we
can verify i) for the relevant vertices by integrating the
coupling constants from
$\Lambda\in[\Lambda_H,\Lambda_0]$ down to $\Lambda_H$ and using Taylor's
formula, as in \REFT; as always, working downwards in $m$ and, for given $m$,
downwards in number of derivatives.  Thus, lemma 6 is verified for the relevant
vertices at order
$r$ in $g$, and therefore for all vertices at this order. By
induction, since it is trivially satisfied at zeroth order in $g$,
the lemma is therefore true at all orders in $g$. \blackbox

\medskip

As far as the proof of Weinberg's theorem is concerned we are only
interested in \eWii\ evaluated at $\Lambda=0$. This tells us that for
$e=0$, so that all momenta and their partial sums have magnitudes greater
than $\Lambda_H$, we have
\eqn\eWxx{ \bigl\Vert \partial^j_p \tilde
G^{c,r}_{2m}(p_1,\ldots,p_{2m};\Lambda_0,\lambda_i) \bigr\Vert_{\Lambda_H} \leq
\Lambda_H^{4-2m-j}P\log\biggl({\Lambda_H \over \Lambda_m}\biggr).}
For all the $p_i$ with magnitudes of order $\Lambda_H$ the damping factors
of $K^{1/4}_{\Lambda_H}(p_i)$ are all of order one and we see that
\eqn\eWxxi{\partial^j_p \tilde
G^{c,r}_{2m}(p_1,\ldots,p_{2m};\Lambda_0,\lambda_i)  \leq
\Lambda_H^{4-2m-j}P\log\biggl({\Lambda_H \over \Lambda_m}\biggr),}
and we have proved Weinberg's theorem to all orders in perturbation
theory for Euclidean momenta with magnitudes
of order $\Lambda_H$, not only for the amputated connected
Green's functions, but also for all their momentum derivatives.

In this section $\Lambda_H$ has no
particular relevance, in the sense that it is not linked to the renormalization
conditions in any way. The only requirement satisfied by $\Lambda_H$
is that it be $\geq m$ and $\leq \Lambda_0$. Thus, we have proved
Weinberg's theorem for all momenta in this range. Also, writing \eWiii\ and
\eWv\ evaluated at $\Lambda=0$ we obtain far more comprehensive bounds
for the magnitudes of Green's functions and their derivatives for
momenta with magnitudes above the mass of the particle, but with
various `exceptional' partial sums with magnitudes below $\Lambda_H$:
for $1\leq e\leq 2m-2$, $m\geq 2$,
\eqn\eWxxxvii{\bigl\Vert \partial^{j}_p
\tilde G^{c}_{2m}(p_1,\ldots,p_{2m};\Lambda_0,\lambda_i)
\bigr\Vert_{\Lambda_H}^{E,e}\leq\cases{
 \Lambda_H^{4-2m-j}
 \Bigl({\Lambda_H\over \bar\Lambda}\Bigr)^{e+j-1}
 P\log \Bigl({\Lambda_H \over \bar\Lambda}\Bigr) & $e$ odd,\cr
 \Lambda_H^{4-2m-j}
 \Bigl({\Lambda_H\over \bar\Lambda}\Bigr)^{e+j-2}
 P\log \Bigl({\Lambda_H \over \bar\Lambda}\Bigr) & $e$ even,\cr}}
where $\bar\Lambda=\max(E,\Lambda_m)$; for $e=2m-1$,
the bound is $\bar\Lambda^{4-2m-j} P\log ({\Lambda_H \over \bar \Lambda})$.
In the same way as described at the end of \S 2.4 we can improve the bounds
on the momentum derivatives by extra factors of
$\bar\Lambda/\Lambda_H$ if some of them act only on unexceptional momenta.

\subsec{Systematic Improvement.}

If we were now to take the conventional view of quantum field theory and take
$\Lambda_0$ to infinity we would recover the conventional form of
Weinberg's theorem. If instead we consider keeping $\Lambda_0$ finite, it is
easy to verify that the bounds \eWxx\ and \eWxxxvii\ are not changed
if we introduce more general boundary conditions on the irrelevant
vertices at $\Lambda_0$, as long as they
satisfy the conditions outlined in \S 3.1 of \REFT.
Indeed, combining the techniques of this section with those in
\S 2.3 and \S 3.1 of \REFT, we can, with a little effort, but no further
insight, prove the result analogous to lemma 4 for Green's functions
with external momenta with magnitudes greater than $\Lambda_R$.
Assuming that all operators with dimension less than or equal to $D$
have couplings which are physically relevant, and thus set on Green's
functions with momenta of order $m$, then for e=0 we can show that
\medskip
\noindent{\it Lemma 7:}
\eqn\eWsia{\Biggl\Vert \biggl({\partial\over\partial\Lambda_0}\partial^j_p
\tilde G^{c}_{2m}(\Lambda_0,\lambda_i)\biggr)_{{\lambda_i}}
\Biggr\Vert_{\Lambda_H} \leq\Lambda_0^{2-D}\Lambda_H^{D+1-2m-j}
P\log\biggl({\Lambda_0\over\Lambda_H}\biggr)P\log\biggl({\Lambda_H\over
\Lambda_m}\biggr),}
and
\eqn\eWsiax{\big\Vert \partial^j_p
\big(\tilde G^{c}_{2m}(\Lambda_0,\lambda_i)
-\hat G^{c}_{2m}(\Lambda_0,\lambda_i)\big)\big\Vert \leq \Lambda_0^{3-D}
\Lambda_H^{D+1-2m-j} P \log \biggl({\Lambda_0 \over
\Lambda_H}\biggr)P\log\biggl({\Lambda_H\over
\Lambda_m}\biggr),}
where $\hat G^{c}_{2m}(\Lambda_0,\lambda_i)$ are the amputated
connected Green's functions for the theory with different boundary
conditions on irrelevant couplings at $\Lambda_0$.
\medskip

The results for $e>0$ are similar: to find the bound on the
$\Lambda_0$--derivative of a theory where we have set low energy
renormalization
conditions on all coupling constants corresponding to operators with
canonical dimension up to $D$, and set all undetermined couplings at
$\lambda_0$ equal to zero, we multiply the right--hand side of \eWxxxvii\ by
$\Lambda_0^{-1}({\Lambda_H\over \Lambda_0})^{D-3}P\log({\Lambda_H\over
\Lambda_m})$; to find the bound on the difference between two
theories, one as described above, the other with the same low energy
renormalization conditions and with undetermined couplings at
$\Lambda_0$ all natural, we multiply the right--hand side of \eWxxxvii\
by $({\Lambda_H\over \Lambda_0})^{D-3}P\log({\Lambda_H\over
\Lambda_m})$. In this we can justify the results quoted in \S 3.2
of \REFT; we may indeed maintain the predictive power of an effective
theory as we consider processes of higher and higher energies by
specifying more and more low energy renormalization conditions.

Since the effective theory makes sense for arbitrarily large external
momenta, we can also consider Green's functions with Euclidean momenta
with magnitudes above $\Lambda_0$, even though these will require the
experimental determination of an infinite number of couplings to
specify them completely. Here however the Weinberg bounds will
generally break down; the best we can do is
use \eWxx\ in the limit $\Lambda_H \to \Lambda_0$ to
show is that for each momentum, and all partial sums, having magnitude
greater than or equal to $\Lambda_0$ then
\eqn\ewsixxx{\partial^j_p \tilde
G^{c,r}_{2m}(p_1,\ldots,p_{2m};\Lambda_0,\lambda_i)  \leq
\Lambda_0^{4-2m-j}P\log({\Lambda_0\over
\Lambda_m})K^{-1/4}_{\Lambda_0}(p_1)\cdots
K^{-1/4}_{\Lambda_0}(p_{2m}):}
the Green's functions are bounded in
the same way as the vertex functions defined at $\Lambda_0$. Clearly
for these very large momenta the Green's
functions should indeed depend strongly on the form of the vertices at
$\Lambda_0$, and we would need more details to provide a
stronger bound than that above. However, it is for momenta with
magnitudes below $\Lambda_0$ that our theory is most `effective',
and we have been able to show that below this naturalness scale Weinberg's
theorem is true for all effective scalar field theories.

By analytically continuing the Euclidean Green's functions we can
also obtain loose bounds on the amputated connected Green's
functions on the boundary
of the physical region. By making a continuation which avoids any
region of exceptional momenta we would naively expect to obtain the bound
\eqn\eWsixac{|\partial^j_p \tilde
G^{c,r}_{2m}(p_1,\ldots,p_{2m};\Lambda_0,\lambda_i)|  \lsim
\Lambda_H^{4-2m-j} P\log ({\Lambda_H \over \Lambda_m}).}
In general this will be correct. However,
the bounds may be violated if we get too close to physical
singularities, such as poles or normal branch points, where the connected
Green's functions may grow very large. Even in these cases there
should be no change in the
dependence on $\Lambda_0$ and the boundary conditions on irrelevant
operators there; this dependence will
simply be the size of the Green's function multiplied by the same
factors of $(\Lambda_H/\Lambda_0)$ and $P\log(\Lambda_H/\Lambda_0)$ as
in \eWsia\ and \eWsiax.

\newsec{Conclusions.}

By exploiting the exact renormalization group we have achieved
two great savings compared to the normal way of investigating infrared
behaviour. Firstly we have not had to worry at all about the
ultraviolet behaviour of the theory when examining the infrared
behaviour\foot{In particular, we note that we have a non--zero mass
term in the bare Lagrangian. Usually arguments for the infrared
behaviour of field theories assume mass counterterms are absent. If
they are present they must be finely tuned in order that the infrared
behaviour for non--exceptional momenta not be spoiled. In our case this
fine tuning is achieved simply by setting the required low energy
renormalization condition on the mass.}, since the two energy scales are
effectively separated. Secondly, even when looking at low energies, we
only have to consider the topological properties of diagrams
containing either one vertex and one loop or two vertices and no
loops, rather than those of arbitrarily complicated graphs. So, as
promised in the introduction, the exact renormalization
group makes the investigation of the scale dependence of Green's
functions much more simple than conventional methods.

One may feel, however, that the description of the infrared behaviour of
a quantum field theory is not particularly interesting for a purely
scalar theory, since the massless case is simply due to a rather
extreme fine-tuning of the renormalization condition on the mass.
However the results of this paper will become very useful when
considering the extension of the methods in \REFT\ to an examination
of the renormalizability of effective gauge
theories in later articles, as well as to the theories containing particles of
significantly different masses discussed in an accompanying paper\Decoupling.

Finally, it is important to realize that we have proven Weinberg's theorem
rather late in the day, and in particular after the boundedness,
convergence, universality, unitarity and causality of the effective
theory have already been established (in \REFT). It was thus clearly
not necessary to know anything of
Weinberg's theorem in order to perform these other proofs. Thus from
the point of view of the exact renormalization group Weinberg's
theorem is no longer to be regarded as an essential ingredient when
proving formal
results in quantum field theory\foot{In fact, we do
not even prove it for individual Feynman diagrams, only for the
complete amputated connected Green's functions.}; rather it should be
regarded as a powerful constraint on the complete
amputated connected Green's functions in the deep Euclidean region, as
well as a good indicator, when combined with the Landau rules, of the
high energy behaviour of S-matrix elements.

\bigskip
\noindent{\bf Acknowledgements.}
\medskip
We would like to thank the Royal Society and the SERC for financial
support during the period when most of this work was done, and the
Theoretical Physics group of Oxford University for their hospitality;
RST would also like to thank Jesus College, Oxford and the
Leathersellers' Company for a graduate scholarship.

\footatend\vfill\supereject\immediate\closeout\rfile\writestoppt
\baselineskip=14pt\centerline{{\bf References}}\bigskip{\frenchspacing%
\parindent=20pt\escapechar=` \input refs.tmp\vfill\eject}\nonfrenchspacing\vfill\eject
\end